\documentclass[12pt]{article}
\usepackage{amssymb,latexsym}
\usepackage[dvips]{graphicx}
\newtheorem{theo}{Theorem}
\newtheorem{defi}[theo]{Definition}

\def\bea{\begin{eqnarray}}
\def\eea{\end{eqnarray}}
\def\nn{\nonumber}
\def\beq{\begin{equation}}
\def\eeq{\end{equation}}

\def\nn{\nonumber}

\def\Z{\mathbb{Z}}
\def\ra{\rangle}
\def\la{\langle}
\def\lb{[\![}
\def\rb{]\!]}

\def\t{\theta}

\renewcommand{\theequation}{\arabic{section}.\arabic{equation}}

\begin{document}
\begin{center}
{\Large \bf A classification of generalized quantum statistics}\\[4mm]
{\Large \bf associated with basic classical Lie superalgebras}\\[3cm]
{\bf N.I.\ Stoilova\footnote{Permanent address:
Institute for Nuclear Research and Nuclear Energy, Boul.\ Tsarigradsko Chaussee 72,
1784 Sofia, Bulgaria} and J.\ Van der Jeugt}\\[2mm]
Department of Applied Mathematics and Computer Science,\\
University of Ghent, Krijgslaan 281-S9, B-9000 Gent, Belgium.\\
E-mails: Neli.Stoilova@UGent.be, Joris.VanderJeugt@UGent.be.
\end{center}


\begin{abstract}
Generalized quantum statistics such as para-statistics is usually characterized
by certain triple relations. In the case of para-Fermi statistics these relations 
can be associated with the orthogonal Lie algebra $B_n=so(2n+1)$; in the case of
para-Bose statistics they are associated with the Lie superalgebra $B(0|n)=osp(1|2n)$.
In a previous paper, a mathematical definition of ``a generalized quantum
statistics associated with a classical Lie algebra $G$'' was given, and a complete
classification was obtained.
Here, we consider the definition of ``a generalized quantum
statistics associated with a basic classical Lie {\em superalgebra} $G$''. 
Just as in the Lie algebra case, this definition is closely related to 
a certain $\Z$-grading of $G$. We give in this paper a complete classification 
of all generalized quantum statistics associated with the basic classical Lie superalgebras 
$A(m|n)$, $B(m|n)$, $C(n)$ and $D(m|n)$. 
\end{abstract}

\vspace{1cm}
\noindent
Running title: Classification of generalized statistics\\[2mm]

\vspace{1cm}

\newpage
\renewcommand{\thesection}{\Roman{section}}
\renewcommand{\theequation}{\arabic{section}.{\arabic{equation}}}

\setcounter{equation}{0}
\section{Introduction} \label{sec:Introduction}%

A historically important extension of Bose and Fermi statistics has been known for 50 years, 
namely the para-Bose and para-Fermi statistics as developed by Green~\cite{Green}. 
In para-statistics, the usual bilinear commutators or anti-commutators for 
bosons and fermions are replaced by certain trilinear or triple relations.
For example, for $n$ pairs of para-Bose creation 
and annihilation operators $B_i^\xi$ ($\xi=\pm$ and $i=1,\ldots,n$), 
the defining relations are:
\begin{eqnarray}
&& [\{B_{ j}^{\xi}, B_{ k}^{\eta}\}, B_{l}^{\epsilon}]=
(\epsilon -\xi) \delta_{jl} B_{k}^{\eta} +  (\epsilon -\eta)
\delta_{kl}B_{j}^{\xi}, \label{para-Bose} \\
&& \qquad\qquad \xi, \eta, \epsilon =\pm\hbox{ or }\pm 1;\quad j,k,l=1,\ldots,n. \nn 
\end{eqnarray}
Similar triple relations hold for the para-Fermi operators $F_i^\xi$~\cite{Green}, see (1.1) in~\cite{GQS}.
Both for para-Bose and para-Fermi statistics, there is a group theoretical setting.
It was shown~\cite{KR} that the Lie algebra generated 
by the $2n$ elements $F^\xi_i$, with $\xi=\pm$ and $i=1,\ldots,n$, 
subject to the para-Fermi relations is $B_n=so(2n+1)$ (as a Lie 
algebra defined by means of generators and relations).

Twenty years after the connection between para-Fermi statistics and the 
Lie algebra $so(2n+1)$, a new connection, between para-Bose statistics 
and the orthosymplectic Lie superalgebra $B(0|n)=osp(1|2n)$~\cite{Kac} 
was discovered~\cite{Ganchev}. 
The Lie superalgebra generated by $2n$ odd 
elements $B^\xi_i$, with $\xi=\pm$ and $i=1,\ldots,n$, subject to the triple
relations~(\ref{para-Bose}), is $osp(1|2n)$ (as a Lie superalgebra 
defined by means of generators and relations). Moreover, there is a certain 
representation of $osp(1|2n)$, the so-called Bose representation ${\cal B}$, 
that yields the classical Bose relations, i.e. where the representatives ${\cal B}(B^\xi_i)$ 
satisfy the relations of classical Bose statistics. For more general 
para-Bose statistics, a class of infinite dimensional $osp(1|2n)$ 
representations needs to be investigated.

These examples show that para-statistics, 
as introduced by Green~\cite{Green} and further developed by others 
(see~\cite{Ohnuki} and  the references therein), 
can be associated with representations of the Lie (super)algebras of class $B$ 
(namely $B_n$ and $B(0,n)$). 
Whether alternative types of generalized quantum statistics can be found
in the framework of other classes of simple Lie algebras or superalgebras has
been considered in particular by Palev~\cite{Palev1}-\cite{sl(1|n)}. 
Building upon his examples and inspired by the definition of creation and annihilation 
operators in~\cite{Palev5}, a mathematical definition of ``generalized quantum
statistics'' was given in~\cite{GQS}. Furthermore, a complete classification was given of 
all the classes of generalized quantum statistics for the classical 
Lie algebras $A_n$, $B_n$, $C_n$ and $D_n$~\cite{GQS}, by means of their algebraic relations.
In the present paper we make a similar classification for the basic
classical Lie superalgebras.

For certain examples of quantum statistics associated with 
 Lie superalgebras, see~\cite{sl(m|n)}-\cite{sl(1|n)}.
However, a complete classification was never made.
A particular interesting example was described for the Lie superalgebra 
$sl(1|n)=A(0|n-1)$~\cite{sl(1|n)}. For this superalgebra, a set of odd creation and annihilation
operators was given~\cite{Palev5}, and it was shown that these $n$ pairs of operators
$a^\xi_i$, with $\xi=\pm$ and $i=1,\ldots,n$, subject to the defining relations
\begin{eqnarray}
&& [\{ a^+_i,a^-_j\},a^+_k]= \delta_{jk} a^+_i - \delta_{ij} a^+_k, \nn\\
&& [\{ a^+_i,a^-_j\},a^-_k]= -\delta_{ik} a^-_j + \delta_{ij} a^-_k, \label{sl1n-relations}\\
&& \{a^+_i,a^+_j\}=\{a^-_i,a^-_j\}=0, \nn
\end{eqnarray}
($i,j,k=1,\ldots,n$), generate the special linear Lie superalgebra $sl(1|n)$ 
(as a Lie superalgebra defined by means of generators and relations). 
Just as in the case of para-Bose relations, (\ref{sl1n-relations}) has two
interpretations. On the one hand, it describes the algebraic
relations of a new kind of generalized statistics, in this case $A$-superstatistics or
a statistics related to the Lie superalgebra $A(0|n-1)$. On the other hand, (\ref{sl1n-relations})
yields a set of defining relations for the Lie superalgebra $A(0|n-1)$ in terms of
generators and relations. Observe that certain microscopic and macroscopic 
properties of this statistics have already been studied~\cite{sl(1|n)}.

A description similar to (\ref{sl1n-relations}) for the Lie algebra $A_n$ 
was given for the first time by N.\ Jacobson~\cite{Jacobson}
 in the context of ``Lie triple systems''. 
Therefore, this type of generators is often referred to as the 
``Jacobson generators''. In this context, we shall mainly
use the terminology ``creation and annihilation operators (CAOs) for $sl(1|n)$''.

Following the mathematical definition of ``generalized
quantum statistics associated with a Lie algebra'', given in~\cite{GQS}, this notion will
be extended to Lie superalgebras $G$. This definition, and the corresponding 
classification method, are described in section~\ref{sec2}. Just as
for the case of Lie algebras, the method leads to a
classification of certain gradings of $G$, and to regular subalgebras of $G$.
In this process, Dynkin diagram techniques play a crucial role. 
For the basic classical Lie superalgebras however, the description by
means of a Dynkin diagram is not unique: besides the so-called distinguished
Dynkin diagram, other non-equivalent Dynkin diagrams exist~\cite{Kac}, \cite{Serganova}.
This feature will make it harder to obtain a complete classification
of all generalized quantum systems. 
In the remaining sections, the classification results are presented for
all basic classical Lie superalgebras. 
A final section discusses some possible applications.

For the basic classical Lie superalgebras~\cite{Kac}, we use the notation $A(m|n)=sl(m+1|n+1)$,
$B(m|n)=osp(2m+1|2n)$, $C(n)=osp(2|2n-2)$ and $D(m|n)=osp(2m|2n)$. The algebra $B(0|n)=osp(1|2n)$
has a different structure and is usually considered separately (also here).
For the classical simple Lie algebras, we use the notation $A_n=sl(n+1)$, $B_n=so(2n+1)$,
$C_n=sp(2n)$ and $D_n=so(2n)$; note the difference between $C_n$ and $C(n)$.
Note also that for trivial values of $m$ or $n$, a Lie superalgebra coincides with
a Lie algebra: $sl(r|0)=sl(0|r)=sl(r)$, $B(m|0)=B_m$, $D(m|0)=D_m$, $D(0|n)=C_n$.

\setcounter{equation}{0}
\section{Definition and classification method} \label{sec2}%

Let $G$ be a basic classical Lie superalgebra. $G$ has a $\Z_2$-grading $G=G_{\bar 0}\oplus G_{\bar 1}$;
an element $x$ of $G_{\bar 0}$ is an even element ($\deg(x)=0$), 
an element $y$ of $G_{\bar 1}$ is an odd element ($\deg(y)=1$). The elements
which are purely even or odd are called homogeneous elements.
The Lie superalgebra bracket is denoted by $\lb x,y \rb$. In the universal enveloping
algebra of $G$, this stands for 
\[
\lb x,y \rb = x y - (-1)^{\deg(x)\deg(y)} y x,
\]
if $x$ and $y$ are homogeneous. So the bracket can be a commutator or an anti-commutator.

Just as for a Lie algebra~\cite{GQS}, a generalized quantum statistics associated with 
$G$ is determined by a set of $N$ creation operators $x_i^+$ and $N$ annihilation
operators $x_i^-$. Following the ideas of para-Bose statistics and those of~\cite{GQS},
these $2N$ operators should generate the Lie superalgebra $G$, subject to certain triple
relations. Let $G_{+1}$ and $G_{-1}$ be the subspaces of $G$ spanned by these elements:
\begin{equation}
G_{+1} = \hbox{span} \{x^+_i;\ i=1\ldots,N\},\qquad
G_{-1} = \hbox{span} \{x^-_i;\ i=1\ldots,N\}.
\end{equation}
We do not require that these subspaces are homogeneous.
The space $\lb G_{+1},G_{+1} \rb$ can be zero (in which case the creation operators
mutually supercommute, as in~(\ref{sl1n-relations})) or non-zero (as in~(\ref{para-Bose})).
A similar statement holds for the annihilation operators and $\lb G_{-1},G_{-1} \rb$.
Putting $G_{\pm 2}=\lb G_{\pm 1},G_{\pm 1}\rb$ and $G_0=\lb G_{+1},G_{-1}\rb$, 
the condition that $G$ is generated by the $2N$ elements subject to triple relations
only, leads~\cite{GQS} to the requirement that $G= G_{-2} \oplus G_{-1} \oplus G_0 \oplus G_{+1} \oplus G_{+2}$,
and this must be a $\Z$-grading of $G$.
Since these subspaces are not necessarily homogeneous, this $\Z$-grading is in general
not consistent with the $\Z_2$-grading.

Just as in~\cite{GQS}, we shall impose two further requirements: first of all,
the generating elements $x_i^\pm$ must be root vectors of $G$.
Secondly, $\omega(x_i^+)=x_i^-$, where $\omega$ is the standard antilinear anti-involutive
mapping of $G$ (in terms of root vectors $e_\alpha$, $\omega$ satisfies
$\omega(e_\alpha)=e_{-\alpha}$).
This leads to the following definition, completely analogous as in~\cite{GQS}:

\begin{defi}
Let $G$ be a basic classical Lie superalgebra, with antilinear anti-involutive mapping $\omega$.
A set of $2N$ root vectors $x^\pm_i$ ($i=1,\ldots,N$) is called a set of
creation and annihilation operators for $G$ if:
\begin{itemize}
\item $\omega(x^\pm_i)=x^\mp_i$,
\item $G= G_{-2} \oplus G_{-1} \oplus G_0 \oplus G_{+1} \oplus G_{+2}$ is
a $\Z$-grading of $G$, with $G_{\pm 1}= \hbox{span}\{x^\pm_i,\ i=1\ldots,N\}$
and $G_{j+k}=\lb G_j,G_k \rb$.
\end{itemize}
The algebraic relations ${\cal R}$ satisfied by the operators $x_i^\pm$
are the relations of a generalized quantum statistics (GQS) associated with $G$.
\end{defi}

This is a {\em mathematical} generalization of quantum statistics. 
Whether all such GQS actually lead to physically acceptable quantum
statistics remains to be seen; in this sense one should interpret our
GQS as ``candidates for generalizations of quantum statistics''. 

A GQS is characterized by a set $\{x_i^\pm\}$ of CAOs
and the set of algebraic relations ${\cal R}$ they satisfy.
A consequence of this definition is that $G$ is generated by $G_{-1}$ and $G_{+1}$,
i.e.\ by the set of CAOs, and since $G_{j+k}=\lb G_j,G_k \rb$, it follows that
\begin{equation}
G=\hbox{span}\{ x_i^\xi,\ \lb x_i^\xi,x_j^\eta \rb; \quad i,j=1,\ldots,N,\ \xi,\eta=\pm\}.
\end{equation} 
This implies that it is necessary and sufficient to give all relations of the following type:
\begin{itemize}
\item[(R1)] The set of all linear relations between the elements $\lb x_i^\xi, x_j^\eta\rb$ 
($\xi,\eta=\pm$, $i,j=1,\ldots,N$). 
\item[(R2)] The set of all triple relations of the form $\lb \lb x_i^\xi, x_j^\eta \rb,x_k^\zeta \rb=
\hbox{linear combination of }x_l^\theta$. 
\end{itemize}
So ${\cal R}$ consists of a set of quadratic relations and a set of triple relations.
Also, as a Lie superalgebra defined by generators and relations, $G$ is uniquely characterized 
by the set of generators $x_i^\pm$ subject to the relations ${\cal R}$.

A consequence of this definition is that $G_0$ itself is a subalgebra of $G$
spanned by root vectors of $G$~\cite{GQS}. It follows that $G_0$ is a regular subalgebra 
containing the Cartan subalgebra $H$ of $G$.
By the adjoint action, the remaining $G_i$'s are $G_0$-modules.
Thus the technique of~\cite{GQS} can be used in order to obtain a complete classification
of all GQS associated with $G$:
\begin{enumerate}
\item
Determine all regular subalgebras $G_0$ of $G$. If not yet contained in $G_0$, replace $G_0$
by $G_0 + H$.
\item
For each regular subalgebra $G_0$, determine the decomposition of $G$
into simple $G_0$-modules $g_k$ ($k=1,2,\ldots$).
\item
Investigate whether there exists a $\Z$-grading of $G$ of the form
\begin{equation}
G=G_{-2} \oplus G_{-1} \oplus G_0 \oplus G_{+1} \oplus G_{+2}, 
\label{5grading}
\end{equation}
where
each $G_i$ is either directly a module $g_k$ or else a sum of
such modules $g_1\oplus g_2\oplus \cdots$, such that
$\omega (G_{+i})=G_{-i}$.
\end{enumerate}

If the $\Z$-grading is of the form~(\ref{5grading}) with $G_{\pm 2}\ne 0$, we shall say
that it has {\em length}~5; if $G_{+2}=0$ (then $G_{-2}=0$, but $G_{\pm1}\ne0$), then the
$\Z$-grading is of length~3.

To find regular subalgebras one can use the method of (extended) Dynkin diagrams~\cite{regular}. 
The second stage is straightforward by means of representation theoretical
techniques.
The third stage requires most of the work: one must try out all possible
combinations of the $G_0$-modules $g_k$, and see whether it is possible to
obtain a grading of the type~(\ref{5grading}). In this process, if one of the 
simple $G_0$-modules $g_k$ is such that $\omega(g_k)=g_k$, then it follows
that this module should be part of $G_0$. In other words, such a case
reduces essentially to another case with a larger regular subalgebra.

In the following sections we shall give a summary of the classification
process for the basic classical Lie superalgebras 
$A(m|n)$, $B(m|n)$, $B(0|n)$, $D(m|n)$ and $C(n)$. 
Note that, in order to identify a GQS associated
with $G$, it is sufficient to give only the set of CAOs, or alternatively, to give
the subspace $G_{-1}$ (then the $x_i^-$ are the root vectors of $G_{-1}$,
and $x_i^+=\omega(x_i^-)$ ). The set ${\cal R}$ then consist of 
all quadratic relations (i.e. the linear relations between the elements
$\lb x_i^\xi, x_j^\eta\rb$) and all triple relations, and all of these
relations follow from the known supercommutation relations in $G$.
Because, in principle, ${\cal R}$ can be determined from the set $\{x_i^\pm; i=1,\ldots,N\}$, we
will not always give it explicitly. 

Finally, observe that two different sets of CAOs $\{x_i^\pm; i=1\ldots,N\}$ and
$\{y_i^\pm; i=1\ldots,N\}$ (same $N$) are said to be isomorphic if, for a certain
permutation $\tau$ of $\{1,2,\ldots,N\}$, the relations between the elements
$x_{\tau(i)}^\pm$ and $y_i^\pm$ are the same. In that case, the regular subalgebra
$G_0$ spanned by $\{\lb x_i^+,x_j^-\rb \}$ is isomorphic (as a Lie superalgebra) to the
regular subalgebra spanned by $\{\lb y_i^+,y_j^-\rb \}$.

\setcounter{equation}{0}
\section{The Lie superalgebra $A(m|n)$} \label{sec:A}%

Let $G$ be the special linear Lie superalgebra $A(m|n)\equiv sl(m+1|n+1)$, consisting of traceless
$(m+n+2)\times (m+n+2)$ matrices. The Cartan subalgebra $H$ of $G$ is the
subspace of diagonal matrices. The root vectors of $G$ are known to be the
elements $e_{jk}$ ($j\ne k=1,\ldots,m+n+2$), where $e_{jk}$ is a matrix with
zeros everywhere except a 1 on the intersection of row~$j$ and column~$k$. The 
$\Z_2$-grading is such that $\deg(e_{jk})=\theta_{jk}=
\theta_j +\theta_k$, where 
\begin{eqnarray}
&&\theta_j= \left\{ \begin{array}{lll}
 {0} & \hbox{if} & j=1,\cdots ,m+1 \\ 
 {1} & \hbox{if} & j=m+2,\cdots ,m+n+2.
 \end{array}\right.
\end{eqnarray}
The root corresponding to $e_{jk}$ ($j,k=1,\ldots,m+1$) is given by $\epsilon_j-\epsilon_k$;
for $e_{m+1+j,m+1+k}$ ($j,k=1,\ldots,n+1$) it is $\delta_j-\delta_k$;
and for $e_{j,m+1+k}$, resp.\ $e_{m+1+k,j}$, ($j=1,\ldots,m+1$, $k=1,\ldots,n+1$) it is $\epsilon_j-\delta_k$,
resp.\ $\delta_k-\epsilon_j$.
The anti-involution is such that $\omega(e_{jk})=e_{kj}$.
The distinguished set of simple roots and the distinguished Dynkin diagram of $A(m|n)$ are given in Table~1,
and so is the extended distinguished Dynkin diagram.

To find regular subalgebras of $G=A(m|n)$, one should delete nodes from the Dynkin diagrams of
$A(m|n)$ (first the ordinary,
and then the extended). This goes in systematic
steps. For each step, we shall investigate whether it leads to a grading
of type~(\ref{5grading}).

\noindent
{\bf Step 1.} Delete node $i$ from the distinguished 
Dynkin diagram. Then the corresponding diagram is 
the Dynkin diagram of $G_0=sl(i)\oplus sl(m+1-i|n+1)$ for $i=1,\ldots,m+1$ and 
of $G_0=sl(m+1|i-m-1)\oplus sl(n+m+2-i)$ for $i=m+2,\ldots,m+n+1$. 
There are only two $G_0$-modules and 
\begin{eqnarray}
&&G_{-1}={\rm span}\{ e_{kl};\ k=1,\ldots ,i,\ l=i+1,\ldots ,m+n+2\},
\label{exA1}\\
&& G_{+1}={\rm span}\{ e_{lk};\ k=1,\ldots ,i,\ l=i+1,\ldots ,m+n+2\}.
\end{eqnarray}
Therefore $A(m|n)$ has a grading of length~3, $A(m|n)=G_{-1}\oplus G_0\oplus G_{+1}$,
and the number of creation and annihilation operators is $N=i(m+n+2-i)$. 

The most interesting realizations are those with $i=1$, $i=m+n+1$,  
$i=2$ and $i=m+n$. We shall give the explicit supercommutation relations between the CAOs
for some of these cases.

For $i=1$, $N=m+n+1$. Putting
\[ 
a_j^-=e_{1,j+1}, \quad a_j^+=e_{j+1,1},\qquad j=1,\cdots ,m+n+1, 
\]
the relations ${\cal R}$ are:
\begin{eqnarray}
&&\lb a_j^+,a_k^+\rb =\lb a_j^-,a_k^-\rb=0,\nn\\
 &&\lb \lb a_j^+,a_k^-\rb ,a_l^+\rb =(-1)^{\t_{j+1}}\delta_{jk}a_l^++\delta_{kl}a_j^+ , \label{A1}
\\
&&\lb \lb a_j^+,a_k^-\rb ,a_l^-\rb =-(-1)^{\t_{j+1}}\delta_{jk}a_l^--
(-1)^{\t_{j+1,k+1}\t_{l+1}}\delta_{jl}a_k^- . \nn
\end{eqnarray}
For $m=0$, these are the relations of $A$-superstatistics~\cite{Palev5}, \cite{sl(1|n)}, see~(\ref{sl1n-relations}).
Also for general $m$ and $n$, these relations have been considered in another context~\cite{sl(m|n)}.

For $i=m+n+1$, $N=m+n+1$. Putting
\[ 
a_j^-=e_{j,m+n+2}, \quad a_j^+=e_{m+n+2,j},\qquad j=1,\cdots ,m+n+1 
\]
one finds:  
\begin{eqnarray}
  &&\lb a_j^+,a_k^+\rb =\lb a_j^-,a_k^-\rb=0, \nn\\
 &&\lb \lb a_j^+,a_k^-\rb ,a_l^+\rb =\delta_{jk}a_l^+-(-1)^{\t_k}\delta_{kl}a_j^+ , \label{A2} 
\\
&&\lb \lb a_j^+,a_k^-\rb ,a_l^-\rb =-\delta_{jk}a_l^--
(-1)^{(\t_{j}+1)(\t_k+1)}\delta_{jl}a_k^- . \nn
\end{eqnarray}
The relations (\ref{A1}) and (\ref{A2}) are similar; however the corresponding GQS are not isomorphic.
For instance, in (\ref{A1}) there are $m$ even and $n+1$ odd pairs of CAOs, and in
(\ref{A2}) there are $n$ even and $m+1$ odd pairs of CAOs.

For $i=2$, $N=2(m+n)$. One puts
\begin{eqnarray}
&& a_{-,j}^-=e_{1,j+2}, \qquad a_{+,j}^-=e_{2,j+2},\qquad j=1,\ldots ,m+n,\nn
\\
&&a_{-,j}^+=e_{j+2,1}, \qquad a_{+,j}^+=e_{j+2,2},\qquad j=1,\ldots ,m+n. \nn
\end{eqnarray}
Then the corresponding relations read ($\xi, \eta, \epsilon =\pm$; $j,k,l=1,\ldots,m+n$):
\begin{eqnarray}
  &&\lb a_{\xi j}^+,a_{\eta k}^+\rb =\lb a_{\xi j}^-,a_{\eta k}^-\rb =0, \nn\\
&& \lb a_{\xi j}^+,a_{-\xi k}^-\rb =0, \qquad j\neq k,  \label{Adouble} \\
&& \lb a_{-j}^+,a_{- k}^-\rb =\lb a_{+j}^+,a_{+k}^-\rb , \qquad j\neq k, \nn \\
&& \lb a_{+j}^+,a_{- j}^-\rb =\lb a_{+k}^+,a_{- k}^-\rb, \qquad\hbox{ for } \t_{j}=\t_{k}, \nn\\
&& \lb a_{-j}^+,a_{+ j}^-\rb =\lb a_{-k}^+,a_{+ k}^-\rb, \qquad\hbox{ for } \t_{j}=\t_{k}, \nn\\
 &&\lb\lb a_{\xi j}^+,a_{\eta k}^-\rb ,a_{\epsilon l}^+\rb=
 (-1)^{\deg(a_{\xi j}^+)\deg(a_{\eta k}^-)+\delta_{\xi,-\eta}\t_{12}
 \deg(a_{\epsilon l}^+)}
 \delta_{\eta\epsilon}\delta_{jk}
 a_{\xi l}^++\delta_{\xi\eta}\delta_{kl}a_{\epsilon j}^+ , \nn
\\
&&\lb \lb a_{\xi j}^+,a_{\eta k}^-\rb ,a_{\epsilon l}^-\rb =-
(-1)^{\deg(a_{\xi j}^+)\deg(a_{\eta k}^-)}
\delta_{\xi \epsilon}\delta_{jk}
 a_{\eta l}^--
 (-1)^{\t_{j+2,k+2}\deg(a_{\epsilon l}^-)}
 \delta_{\xi\eta}\delta_{jl}a_{\epsilon k}^- . \nn
\end{eqnarray}
Such relations are definitely more complicated than (\ref{A1}) or (\ref{A2}). However, they
are still proper defining relations for $A(m|n)$.

\smallskip \noindent {\bf Step 2.} 
Delete node $i$ and $j$ from the distinguished Dynkin diagram.
We have $G_0=H+sl(i)\oplus sl(j-i)\oplus sl(m+1-j|n+1)$ for
$1\leq i<j\leq m+1$, $G_0=H+sl(i)\oplus sl(m+1-i|j-m-1)\oplus sl(m+n+2-j)$ for
$1\leq i\leq m+1$, $m+2\leq j\leq m+n+1$ and 
$G_0=H+sl(m+1|i-m-1)\oplus sl(j-i)\oplus sl(m+n+2-j)$ for $m+2\leq i<j\leq m+n+1$. 
There are six simple $G_0$-modules. All the possible combinations 
of these modules give rise to gradings of length~5.
There are essentially three different ways in which these $G_0$-modules
can be combined. To characterize these three cases, it is sufficient
to give only $G_{-1}$:
\begin{eqnarray}
G_{-1}&=&\hbox{span}\{ e_{kl},e_{lp};\ k=1,\ldots ,i,
\ l=i+1,\ldots ,j,\ p=j+1, \ldots, m+n+2\},\nn\\
&& \hbox{with }N=(j-i)(m+n+2-j+i); \label{A21}\\
G_{-1}&=&\hbox{span}\{ e_{kl},e_{pk};\ k=1,\ldots ,i,
\ l=i+1,\ldots ,j,\ p=j+1, \ldots, m+n+2\}, \nn \\
&& \hbox{with }N=i(m+n+2-i); \label{A22}\\
G_{-1}&=&\hbox{span}\{ e_{kl},e_{lp};\ k=1,\ldots ,i,
\ p=i+1,\ldots ,j,\ l=j+1, \ldots, m+n+2\}, \nn \\
&& \hbox{with }N=j(m+n+2-j).\label{A23} 
\end{eqnarray}
Note that a part of the solutions in~(\ref{A22}) and~(\ref{A23})
are isomorphic to some of those given by~(\ref{A21}). The isomorphic 
cases can be recognized as those having the same Dynkin diagram of $G_0$
and the same $N$-value.

For reasons explained earlier, we shall no longer give the
corresponding set of relations explicitly for all possible cases.
As an example, we consider here the case $j-i=1$ and~(\ref{A21}).
Then there are $N=m+n+1$ pairs of CAOs, which we can label as follows:
\begin{eqnarray*}
&&a_k^-=e_{k,i+1}, \quad a_k^+=e_{i+1,k}, \qquad k=1,\ldots ,i;\\ 
&&a_k^-=e_{i+1,k+1}, \quad a_k^+=e_{k+1,i+1}, \qquad k=i+1,\ldots ,m+n+1.
\end{eqnarray*}
Using 
\begin{equation}
\langle k\rangle= \left\{ \begin{array}{lll}
 {0} & \hbox{if} & k=1,\ldots ,i, \\ 
 {1} & \hbox{if} & k=i+1,\ldots ,m+n+1,
 \end{array}\right.
\label{twokinds} 
\end{equation}
the quadratic and triple relations now read:
\begin{eqnarray}
 &&\lb a_k^+,a_l^+\rb =\lb a_k^-,a_l^-\rb =0, \qquad k,l=1,\ldots, i \ \hbox{or}\
 k,l =i+1, \ldots ,m+n+1,\nn\\
 && \lb a_k^-,a_l^+\rb =\lb a_k^+,a_l^-\rb =0, \qquad k=1,\ldots , i, \ l=i+1,\ldots, m+n+1,\nn \\
 &&\lb \lb a_k^+,a_l^-\rb ,a_p^+\rb =(-1)^{\la l\ra +\la p \ra +\la k \ra 
 \t_{k+1,i+1}}\delta_{kl}a_p^++
 (-1)^{\la l\ra +\la p \ra +(1-\la l \ra )\t_{l,i+1}(\t_{lk}+\t_{k,i+1})} 
 \delta_{lp}a_k^+ ,\nn\\
 &&
 \qquad \qquad \qquad \qquad \qquad\qquad \ \  k,l=1,\cdots, i,\ {\rm or} \ k,l=i+1,\ldots , m+n+1, \nn \\
&&\lb \lb a_k^+,a_l^-\rb ,a_p^-\rb =
-(-1)^{\la l\ra 
+\la p \ra + \deg(a_k^+)[\la k \ra \t_{k+1,l+1}+(1-\la l \ra)\t_{l,i+1}]}
\delta_{kp}a_l^-\nn\\
&&
-(-1)^{\la l\ra +\la p \ra +\la k \ra \t_{k+1,i+1}}\delta_{kl}a_p^-,
  \qquad k,l=1,\cdots, i,\ {\rm or} \ k,l=i+1,\ldots , m+n+1, \nn \\
&&\lb \lb a_k^{\xi},a_l^{\xi}\rb ,a_p^{-\xi}\rb =
-(-1)^{{\frac{1} {2}}\t_{p,i+1}[(1+\xi)\t_{l+1,i+1}+
(1-\xi)\t_{k,l+1}]}\delta_{kp}a_l^{\xi}\nn \\
&&
+(-1)^{{\frac{1} {2}}(1+\xi)\t_{l+1,i+1}(\t_{k,i+1}+\t_{k,l+1})}
\delta_{lp}a_k^{\xi}, \qquad k=1,\ldots, i, \ l=i+1,\ldots ,m+n+1 ,\nn \\
&&\lb \lb a_k^{\xi},a_l^{\xi}\rb ,a_p^{\xi}\rb =0 ,\qquad 
\xi=\pm;\ k,l,p=1,\ldots,m+n+1.\label{A21R}  
\end{eqnarray}

\smallskip \noindent {\bf Step 3.} 
If we delete three or more nodes from the distinguished Dynkin diagram, the resulting
$\Z$-gradings of $A(m|n)$ are no longer of the form (\ref{5grading}).
So these cases are not relevant for our classification.

\smallskip \noindent {\bf Step 4.} 
Next, we move on to the extended distinguished Dynkin diagram, also given in Table~1.
If we delete node $i$ from this extended diagram,
the remaining diagram is again (a non-distinguished Dynkin diagram) of type $A(m|n)$, so $G_0=G$,
and there are no CAOs.

\smallskip\noindent {\bf Step 5.} 
If we delete node $i$ and $j$ ($i<j$) from the extended distinguished Dynkin diagram,
then $A(m|n)=G_{-1}\oplus G_0\oplus G_{+1}$ with $G_0=H+sl(m|n+1)$
or $H+sl(m+1|n)$ when the nodes are adjacent, and $G_0=H+sl(k|l)\oplus sl(p|q)$
with $k+p=m+1$ and $l+q=n+1$ when the nodes are nonadjacent. Note that $p$
or $q$ can be zero: $sl(r|0)=sl(0|r)=sl(r)$. Now
\begin{equation}
G_{-1}=\hbox{span}\{ e_{kl};\ k=i+1\ldots ,j,\ l\neq i+1,\ldots ,j\}.
\label{exA2}
\end{equation}
The number of annihilation operators is $N=(j-i)(n+m+2-j+i)$.
A part of these solutions are isomorphic to some of those of Step~1.
The isomorphic cases are again characterized by the fact that their $G_0$'s are 
isomorphic Lie superalgebras and their $N$-values coincide.

\smallskip\noindent {\bf Step 6.} 
If we delete nodes $i$, $j$ and $k$ from the extended distinguished 
Dynkin diagram ($i<j<k$), then the corresponding $\Z$-gradings are of the form~(\ref{5grading}).
If the three nodes are adjacent $G_0=H+sl(m-1|n+1)$, $H+sl(m|n)$ or $H+sl(m+1|n-1)$.
When two adjacent and one nonadjacent nodes are deleted, 
$G_0=H+sl(l|p)\oplus sl(q|r)$ with $l+q=m$, $p+r=n+1$ or
$l+q=m+1$, $p+r=n$. 
If all three nodes are nonadjacent  then
$G_0=H+sl(l|p)\oplus sl(q|r)\oplus sl(s|t)$ with $l+q+s=m+1$, 
$p+r+t=n+1$. One or two of these three Lie superalgebras is 
$sl(r|0)=sl(0|r)=sl(r)$.
There are three different ways in which the corresponding $G_0$-modules
can be combined. We give here only $G_{-1}$:
\begin{eqnarray}
G_{-1}&=&\hbox{span}\{ e_{ps},e_{sq};\ p=1,\ldots ,i,k+1,\ldots ,n+m+2, \ 
s=i+1,\ldots ,j,\ q=j+1, \ldots, k\},\nn\\
&& \hbox{with }N=(j-i)(n+m+2-j+i); \label{exA3}\\
G_{-1}&=&\hbox{span}\{ e_{ps},e_{qp};\ p=1,\ldots ,i,k+1,\ldots ,n+m+2,
\ s=i+1,\ldots ,j,\ q=j+1, \ldots, k\}, \nn\\
&& \hbox{with }N=(k-i)(n+m+2+i-k); \label{exA4}\\
G_{-1}&=&\hbox{span}\{ e_{pq},e_{qs};\ p=1,\ldots ,i,k+1,\ldots ,n+m+2,
\ s=i+1,\ldots ,j,\ q=j+1, \ldots, k\}, \nn\\
&& \hbox{with }N=(k-j)(n+m+2+j-k).\label{exA5} 
\end{eqnarray}
Again a part of these solutions are isomorphic to some of those in Step~2
(characterized by an isomorphic $G_0$ and the same $N$).

\smallskip\noindent {\bf Step 7.} 
If we delete four or more nodes from the extended distinguished 
Dynkin diagram, the corresponding
$\Z$-grading of $A(m|n)$ has no longer the required properties (i.e.\ there are
non-zero subspaces $G_i$ with $|i|>2$).

\smallskip\noindent {\bf Step 8.} 
Next, one should repeat the process for all non-distinguished Dynkin
diagrams of $G$ and their extensions. This is what makes the work harder
than the corresponding classification for Lie algebras (which have only
one Dynkin diagram and one extension).
A general Dynkin diagram is determined by a general simple root
system. All the systems of simple roots $\Pi_{S,T}$ of $A(m|n)$ 
are determined by two increasing sequences~\cite{Kac}, \cite{Serganova}
\[
S=\{ s_1<s_2<\ldots \} \ {\rm and} \ 
T=\{ t_1<t_2<\ldots \} 
\]
and a sign:
\[
\Pi_{S,T}=\pm \{ \epsilon_1-\epsilon_2, \epsilon_2-\epsilon_3,\ldots,
\epsilon_{s_1}-\delta_1, \delta_1-\delta_2,\ldots, \delta_{t_1}-\epsilon_{s_1+1}, \ldots \}
\]
The Dynkin diagram itself looks like
\[
\includegraphics{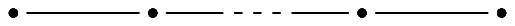}
\]
where each dot can be a white or gray circle (depending upon whether the corresponding
simple root is even or odd).
Let $\pm (\eta_i-\eta'_j)$ be the last element of $\Pi_{S,T}$ ($\eta$ and $\eta'$ can
be $\epsilon$ or $\delta$). Then the zero node of the extended Dynkin diagram corresponds to 
$\pm (\eta'_j-\epsilon_1)$ and it is uniquely determined.

If we repeat the whole procedure with the non-distinguished Dynkin diagrams (ordinary
and extended), the only new result corresponds to Step~6 deleting three nonadjacent nodes from
the extended Dynkin diagram. We have $G_0=H+sl(l|p)\oplus sl(q|r)\oplus sl(s|t)$ with $l+q+s=m+1$, 
$p+r+t=n+1$ and in some cases none of the three algebras is 
$sl(r|0)=sl(0|r)=sl(r)$. Just as in Step~2 or~6, there are three different ways in which the
$G_0$-modules can be combined; the explicit expression is left to the reader.

\section{The Lie superalgebras $B(m|n)$}
\setcounter{equation}{0} \label{sec:B}

$G=B(m|n)\equiv osp(2m+1|2n)$ is the subalgebra of $sl(2m+1|2n)$ consisting of matrices of the form:
\begin{equation}
\left(\begin{array}{ccccc} a&b&u&x&x_1  \\
c&-a^t&v&y&y_1\\
-v^t&-u^t&0&z&z_1\\
y_1^t&x_1^t&z_1^t&d&e\\
-y^t&-x^t&-z^t&f&-d^t
\end{array}\right),
\label{osp}
\end{equation}
where $a$ is any $(m\times m)$-matrix, $b$ and $c$ are antisymmetric $(m\times m)$-matrices, 
$u$ and $v$ are $(m\times 1)$-matrices, $x,y,x_1,y_1$ are $(m\times n)$-matrices, $z$ and
$z_1$ are $(1\times n)$-matrices, $d$ is any $(n\times n)$-matrix,
and $e$ and $f$ are symmetric  $(n\times n)$-matrices. The even elements have 
$x=y=x_1=y_1=0$, $z=z_1=0$ and the odd elements are those with 
$a=b=c=0$, $u=v=0$, $d=e=f=0$. We shall consider $m=0$ separately in the next section. 
The Cartan subalgebra $H$ of $G$ is again the
subspace of diagonal matrices $D$. 
Putting $\epsilon_i(D)=D_{ii}, i=1,\ldots, m$ and
$\delta_i(D)=D_{2m+i+1,2m+i+1}, i=1,\ldots,n$, the even 
root vectors and corresponding roots of $G$ are given by:
\begin{eqnarray*}
e_{jk}-e_{k+m,j+m} & \leftrightarrow &\epsilon_j -\epsilon_k, \qquad j\neq k=1,\ldots ,m,\\
e_{j,k+m}-e_{k,j+m} & \leftrightarrow &\epsilon_j +\epsilon_k, \qquad j<k=1,\ldots ,m,\\
e_{j+m,k}-e_{k+m,j} & \leftrightarrow &-\epsilon_j -\epsilon_k, \qquad j<k=1,\ldots ,m,\\
e_{j,2m+1}-e_{2m+1,j+m} & \leftrightarrow &\epsilon_j,  \qquad j=1,\ldots ,m,\\
e_{j+m,2m+1}-e_{2m+1,j} & \leftrightarrow & -\epsilon_j , \qquad j=1,\ldots ,m,\\
e_{2m+1+j,2m+1+k}-e_{n+2m+1+k,n+2m+1+j} & \leftrightarrow &\delta_j -\delta_k, 
\qquad j\neq k=1,\ldots ,n,\\
e_{2m+1+j,2m+1+k+n}+e_{2m+1+k,2m+1+j+n} & \leftrightarrow 
&\delta_j +\delta_k, \qquad j\leq k=1,\ldots ,n,\\
e_{2m+1+n+j,2m+1+k}+e_{2m+1+n+k,2m+1+j} & \leftrightarrow &-
\delta_j -\delta_k, \qquad j\leq k=1,\ldots ,n,
\end{eqnarray*}
and the odd ones by
\begin{eqnarray*}
e_{j,2m+1+k}-e_{2m+1+n+k,j+m} & \leftrightarrow &\epsilon_j -\delta_k, 
\qquad j=1,\ldots ,m, \ k=1,\ldots ,n,\\
e_{m+j,2m+1+k}-e_{2m+1+n+k,j} & \leftrightarrow &-\epsilon_j -\delta_k, 
\qquad j=1,\ldots ,m, \ k=1,\ldots , n,\\
e_{2m+1,2m+1+k}-e_{2m+1+n+k,2m+1} 
& \leftrightarrow &-\delta_k, \qquad k=1,\ldots ,n,\\
e_{j,2m+1+n+k}+e_{2m+1+k,m+j} & \leftrightarrow &\epsilon_j +\delta_k, 
\qquad j=1,\ldots ,m, \ k=1,\ldots, n,\\
e_{m+j,2m+1+n+k}+e_{2m+1+k,j} & \leftrightarrow &-\epsilon_j +\delta_k, 
\qquad j=1,\ldots ,m, \ k=1,\ldots, n,\\
e_{2m+1,2m+1+n+k}+e_{2m+1+k,2m+1} & \leftrightarrow &\delta_k, \qquad k=1,\ldots ,n.
\end{eqnarray*}

The distinguished set of simple roots and the corresponding Dynkin diagram
of $B(m|n)$ are given in Table~1.

\smallskip \noindent {\bf Step 1.} 
Delete node $i$ from the distinguished Dynkin diagram. The corresponding diagram is 
the Dynkin diagram of $G_0=H+sl(i)\oplus B(m|n-i)$ for $i=1,\ldots ,n$
and of  $G_0=H+sl(j|n)\oplus B_{m-j}$ for $i=n+j$, $j=1,\ldots ,m$.
There are four simple $G_0$-modules and $B(m|n)=G_{-2}\oplus G_{-1}\oplus G_0\oplus G_{+1}\oplus G_{+2}$,
where for  $i=1,\ldots n$:
\begin{eqnarray}
G_{-1}&=&\hbox{span}\{ e_{2m+1,2m+1+n+k}+e_{2m+1+k,2m+1}, e_{2m+1+k,2m+1+n+l}+e_{2m+1+l,2m+1+n+k},
\nn \\
&&\ e_{2m+1+k,2m+1+l}-e_{n+2m+1+l,n+2m+1+k}, e_{p,2m+1+n+k}+e_{2m+1+k,m+p},
\nn \\
&& \ e_{m+p,2m+1+n+k}+e_{2m+1+k,p};
\ k=1,\ldots ,i,\ l=i+1,\ldots ,n,\ p=1, \ldots, m\},\nn \\
&& \hbox{with }N=2i(m+n)-i(2i-1); \label{B21}
\end{eqnarray} 
and for $i=n+j, j=1,\ldots m$: 
\begin{eqnarray}
G_{-1}&=&\hbox{span}\{ e_{2m+1,2m+1+n+k}+e_{2m+1+k,2m+1}, e_{p,2m+1}-e_{2m+1,m+p}, \nn\\
&& e_{p,m+q}-e_{q,m+p},  e_{pq}-e_{m+q,m+p},  e_{q,2m+1+n+k}+e_{2m+1+k,m+q}, \nn\\
&& e_{m+q,2m+1+n+k}+e_{2m+1+k,q}; \ k=1,\ldots ,n,\ p=1,\ldots ,j,\ q=j+1, \ldots, m\},\nn \\
&& \hbox{with }N=2i(m+n)-i(2i-1). \label{B22}
\end{eqnarray}
It is interesting to give ${\cal R}$ for $i=n+m$, 
because then the number of creation or annihilation operators is $N=n+m$.
One can label (and rescale) the CAOs as follows:
\begin{eqnarray}
&&b_{j}^-\equiv B_{j}^-= -\sqrt{2}(e_{2m+1, 2m+1+n+j}+e_{2m+1+j,2m+1}), 
\quad j=1,\ldots , n, \nn \\
&&b_{j}^+\equiv B_{j}^+= \sqrt{2}(e_{2m+1, 2m+1+j}-e_{2m+1+n+j,2m+1}), 
\quad j=1,\ldots , n,\\ 
&&b_{n+j}^-\equiv F_{j}^-= \sqrt{2}(e_{j, 2m+1}-e_{2m+1,m+j}), 
\quad j=1,\ldots , m, \nn \\ 
&& b_{n+j}^+\equiv  F_{j}^+=\sqrt{2}(e_{2m+1,j}-e_{m+j,2m+1}), \quad j=1,\ldots , m. 
\end{eqnarray}
Note that
\begin{equation}
\deg(b_j^\pm)=\langle j\rangle= \left\{ \begin{array}{lll}
 {1} & \hbox{if} & j=1,\ldots ,n \\ 
 {0} & \hbox{if} & j=n+1,\ldots ,n+m.
 \end{array}\right.
\end{equation}
There are no quadratic relations, and ${\cal R}$ consists of triple relations only:
\begin{eqnarray}
&& \lb\lb b_{ j}^{\xi}, b_{ k}^{\eta}\rb , b_{l}^{\epsilon}\rb =-2
\delta_{jl}\delta_{\epsilon, -\xi}\epsilon^{\langle l \rangle} 
(-1)^{\langle k \rangle \langle l \rangle }
b_{k}^{\eta} +2  \epsilon^{\langle l \rangle }
\delta_{kl}\delta_{\epsilon, -\eta}
b_{j}^{\xi},  \label{GB3}\\
&& \qquad\qquad \xi, \eta, \epsilon =\pm\hbox{ or }\pm 1;\quad j,k,l=1,\ldots,n+m. \nn 
\end{eqnarray}
Note that $B_j^\pm, j=1,\ldots, n$ (resp.\ $F_k^\pm, k=1,\ldots,m$) are para-Bose
(resp.\ para-Fermi) CAOs, namely
\begin{eqnarray}
&& [\{ B_{ j}^{\xi}, B_{ k}^{\eta}\} , B_{l}^{\epsilon}]=
(\epsilon -\xi)
\delta_{jl} B_{k}^{\eta} +  (\epsilon -\eta)
\delta_{kl}B_{j}^{\xi},  \label{pBose} \\
&& \qquad\qquad \xi, \eta, \epsilon =\pm\hbox{ or }\pm 1;\quad j,k,l=1,\ldots,n; \nn \\[1mm]
&& [[F_{ j}^{\xi}, F_{ k}^{\eta}], F_{l}^{\epsilon}]=\frac 1 2
(\epsilon -\eta)^2
\delta_{kl} F_{j}^{\xi} -\frac 1 2  (\epsilon -\xi)^2
\delta_{jl}F_{k}^{\eta},  \label{pFermi} \\
&& \qquad\qquad \xi, \eta, \epsilon =\pm\hbox{ or }\pm 1;\quad j,k,l=1,\ldots,m. \nn 
\end{eqnarray}
The fact that $B(m|n)$ can be generated by $n$ pairs of para-Bose and $m$ pairs of
para-Fermi operators has been discoved in~\cite{Posp}.

\smallskip\noindent {\bf Step 2.} 
If we delete two or more nodes from the distinguished Dynkin diagram, the resulting
$\Z$-gradings of $B(m|n)$ are no longer of the form~(\ref{5grading}).

\smallskip \noindent {\bf Step 3.}
Now we continue with the extended Dynkin diagram, also given in Table~1.
Delete node $i=0,1,\ldots, n+m$ from the extended distinguished Dynkin diagram.
The remaining diagram is that of $G_0=B(m|n)$, $A_1\oplus B(m|n-1)$,
$C_2\oplus B(m,n-2)$, $\ldots$, $C_{n-1}\oplus B(m|1)$,
$C_n\oplus B_m$, $C(n+1)\oplus B_{m-1}$, $D(2|n)\oplus B_{m-2}$, $\dots$,
$D(m-1|n)\oplus A_1$, $D(m|n)$. 
In all these cases there is only one 
$G_0$-module, so there are no contributions to the classification.

\smallskip \noindent {\bf Step 4.} 
Delete the adjacent nodes $(i-1)$ and $i$, $i=2,3,\dots,n$ from the 
extended distinguished Dynkin diagram. The remaining diagram is that of 
$\tilde{G}_0=A_1\oplus B(m|n-2)$ for $i=2$, of
$\tilde{G}_0=C_{i-1}\oplus B(m|n-i)$ for $i=3,\ldots,n-1$ and of
$\tilde{G}_0=C_{n-1}\oplus B_m$ for $i=n$. In each case there are 
seven $\tilde{G}_0$-modules $g_k$, one of which is invariant under $\omega$
(say $g_1$). Then one has to put $G_0=H+\tilde{G}_0+g_1$, and in each 
case one finds $G_0\equiv H+B(m|n-1)$. Now there are four $G_0$-modules and
$B(m|n)=G_{-2}\oplus G_{-1}\oplus G_0\oplus G_{+1} \oplus G_{+2}$ with 
\begin{eqnarray}
G_{-1}&=&\hbox{span}\{ e_{2m+1, 2m+1+n+i}+e_{2m+1+i,2m+1}, e_{2m+1+i,2m+1+n+j}+e_{2m+1+j,2m+1+n+i},
\nn\\
&& e_{2m+1+i, 2m+1+j}-e_{n+2m+1+j,n+2m+1+i}, \ i\neq j =1,\ldots,n; \nn\\
&&  e_{k,2m+1+n+i}+e_{2m+1+i,m+k}, e_{m+k,2m+1+n+i}+e_{2m+1+i,k}, 
\ k=1,\ldots,m \},\nn
\end{eqnarray}
and $N=2(m+n)-1$. 
Observe that in this case $G_{-2}=\hbox{span}\{ e_{2m+1+i, 2m+1+n+i}\}$, and
all these cases are isomorphic to that of Step~1 with $i=1$.

\smallskip \noindent {\bf Step 5.} 
Delete the adjacent nodes $(i-1)$ and $i$, $i=n+j$, $j=1,2,\dots,m$ from the 
extended distinguished Dynkin diagram. The remaining diagram is that of 
$\tilde{G}_0=C_n\oplus B_{m-1}$ for $j=1$, of
$\tilde{G}_0=C(n+1)\oplus B_{m-2}$ for $j=2$,
of $\tilde{G}_0=D(j-1|n)\oplus B_{m-j}$ for $j=3,\ldots,m-2$,
of $\tilde{G}_0=D(m-2|n)\oplus A_{1}$ for $j=m-1$ and of
$\tilde{G}_0=D(m-1|n)$ for $j=m$. In each case there are 
five $\tilde{G}_0$-modules $g_k$, one of which is invariant under $\omega$
(say $g_1$). Then one has to put $G_0=H+\tilde{G}_0+g_1$, and in each 
case one finds $G_0\equiv H+B(m-1|n)$. Now there are two $G_0$-modules and
$B(m|n)=G_{-1}\oplus G_0\oplus G_{+1}$ with 
\begin{eqnarray}
G_{-1}&=&\hbox{span}\{ e_{j, 2m+1}-e_{2m+1,j+m}, e_{j,k+m}-e_{k,j+m}, e_{jk}-e_{k+m,j+m}, 
\ j\neq k =1,\ldots,m; \nn\\
&&  e_{j,2m+1+n+l}+e_{2m+1+l,m+j},  e_{j,2m+1+l}-e_{2m+1+n+l,j+m}, 
\ l=1,\ldots,n \},\label{B5}
\end{eqnarray}
and $N=2(m+n)-1$. All these cases are mutually isomorphic.

\smallskip \noindent {\bf Step 6.} 
Delete the nonadjacent nodes $i,j$, $i<j-1$,  $j=3,\ldots,n$
from the extended distinguished Dynkin diagram.
The remaining diagram is that of 
$\tilde{G}_0=C_i\oplus sl(j-i)\oplus B(m|n-j)$ (for $i=1$
instead of $C_i$ we have $A_1$).
In each case there are 
seven $\tilde{G}_0$-modules $g_k$, one of which is invariant under $\omega$
(say $g_1$). Then one has to put $G_0=H+\tilde{G}_0+g_1$, and in each 
case one finds $G_0\equiv H+sl(j-i)\oplus B(m|n-j+i)$. 
Now there are four $G_0$-modules and $B(m|n)=G_{-2}\oplus G_{-1}\oplus G_0\oplus G_{+1}\oplus G_{+2}$. 
All these cases are isomorphic to those in Step~1 with $i=2,\ldots,n-1$.

\smallskip \noindent {\bf Step 7.} 
Delete the nonadjacent nodes $i,j$, $i=1,\ldots,n$, $j=n+1,\ldots,n+m$
[but $(i,j)\ne (n,n+1)$] from the extended distinguished Dynkin diagram.
The remaining diagram is that of 
$\tilde{G}_0=C_i\oplus sl(j-n|n-i)\oplus B_{n+m-j}$ (for $i=1$
instead of $C_i$ we have $A_1$).
In each case there are 
seven $\tilde{G}_0$-modules $g_k$, one of which is invariant under $\omega$
(say $g_1$). Then one has to put $G_0=H+\tilde{G}_0+g_1$, and 
one always finds $G_0\equiv H+sl(j-n|n-i)\oplus B(n+m-j|i)$. 
Now there are four $G_0$-modules and $B(m|n)=G_{-2}\oplus G_{-1}\oplus G_0\oplus G_{+1}\oplus G_{+2}$,
where
\begin{eqnarray}
G_{-1}&=&\hbox{span}\{ e_{k, 2m+1}-e_{2m+1,k+m}, e_{2m+1,2m+1+n+l}+e_{2m+1+l,2m+1},
\nn\\
&& e_{k,2m+1+n+p}+e_{2m+1+p,m+k},  e_{k,2m+1+p}-e_{2m+1+n+p,m+k},
\nn\\
&&  e_{2m+1+l,2m+1+n+p}+e_{2m+1+p,2m+1+n+l}, e_{2m+1+l,2m+1+p}-e_{2m+1+n+p,2m+1+n+l},
\nn\\
&&  e_{s,2m+1+n+l}+e_{2m+1+l,m+s}, e_{m+s,2m+1+n+l}+e_{2m+1+l,s},
\nn\\
&& k=1,\ldots,j-n,\ l=i+1,\ldots,n,\ p=1,\ldots,i,\ s=j+1-n,\ldots, m\}, \label{B7}
\end{eqnarray}
with $N=2(j-i)(m+n)-(j-i)(2(j-i)-1)$.
All these cases are new (i.e.\ not isomorphic to an earlier case).

\smallskip \noindent {\bf Step 8.} 
Delete the nonadjacent nodes $i,j$, $i<j-1$, $i=n+1,\ldots,n+m-2$,
$j=n+3,n+4,\ldots, n+m$ from the extended distinguished 
Dynkin diagram.
The remaining diagram is that of 
$\tilde{G}_0=D(i-n|n)\oplus sl(j-i)\oplus B_{n+m-j}$ (for $i=n+1$
instead of $D(i-n|n)$ we have $C(n+1)$).
In each case there are 
seven $\tilde{G}_0$-modules $g_k$, one of which is invariant under $\omega$
(say $g_1$). Then one has to put $G_0=H+\tilde{G}_0+g_1$, and 
one finds $G_0\equiv H+sl(j-i)\oplus B(m-j+i|n)$. 
Now there are four $G_0$-modules, and $B(m|n)=G_{-2}\oplus G_{-1}\oplus G_0\oplus G_{+1}\oplus G_{+2}$,
where
\begin{eqnarray}
G_{-1}&=&\hbox{span}\{ e_{k, 2m+1}-e_{2m+1,k+m},  e_{k,m+l}-e_{l,m+k},  e_{kl}-e_{m+l,m+k}, 
\nn\\
&&  e_{k,2m+1+n+p}+e_{2m+1+p,m+k},  e_{k,2m+1+p}-e_{2m+1+n+p,m+k},
\nn\\
&& k=i-n+1,i-n+2,\ldots,j-n, \nn\\
&& l=1,2,\ldots,i-n,j-n+1,j-n+2,\ldots,m,\
p=1,2,\ldots,n \}, \label{B8}
\end{eqnarray}
with $N=2(j-i)(m+n)-(j-i)(2(j-i)-1)$. All these cases are isomorphic to cases of Step~7. 

\smallskip \noindent {\bf Step 9.}
If we delete three or more nodes from the extended distinguished 
Dynkin diagram, the corresponding
$\Z$-grading of $B(m|n)$ has no longer the required properties (i.e.\ there are
non-zero subspaces $G_i$ with $|i|>2$).

The next step consists of repeating this procedure for the non-distinguished Dynkin
diagrams and their extensions. Following~\cite{Serganova} one can obtain all such Dynkin diagrams of $B(m|n)$. 
We have repeated this procedure for all of them, leading to a lot of case studies but
not leading to any new results (i.e.\ each case is isomorphic to one described 
already by means of the distinguished diagram).

\section{The Lie superalgebras $B(0|n)$}
\setcounter{equation}{0} \label{sec:B0}

We consider the Lie superalgebra $B(0|n)$ separately because the distinguished choice 
of the simple roots for $B(0|n)$ is different than that of $B(m|n)$. In Table~1
the distinguished simple roots, the distinguished Dynkin diagram and the
extended distinguished Dynkin diagram are given.

\smallskip \noindent {\bf Step 1.} 
Delete node $i$, $i=1,\ldots,n$ from the distinguished Dynkin diagram. 
The corresponding diagram is the Dynkin diagram of $G_0=H+sl(i)\oplus B(0|n-i)$.
There are four simple $G_0$-modules and $B(0|n)=G_{-2}\oplus G_{-1}\oplus G_0\oplus G_{+1}\oplus G_{+2}$,
where 
\begin{eqnarray}
G_{-1}&=&\hbox{span}\{ e_{1,1+n+k}+e_{1+k,2m+1}, e_{1+k,1+n+l}+e_{1+l,1+n+k}, e_{1+k,1+l}-e_{n+1+l,n+1+k},
\nn \\
&& k=1,\ldots ,i,\ l=i+1,\ldots ,n, \}, \label{B021}
\end{eqnarray}
with $N=2i(n-i)+i$.
We give ${\cal R}$ for $i=n$, 
because then the number of creation or annihilation operators is $N=n$.
One can label the CAOs as follows:
\begin{eqnarray}
&&B_{j}^-= -\sqrt{2}(e_{1, 1+n+j}+e_{1+j,1}), 
\quad j=1,\ldots , n,\nn\\
&&B_{j}^+= \sqrt{2}(e_{1, 1+j}-e_{1+n+j,1}), 
\quad j=1,\ldots , n.
\end{eqnarray}
These are all odd generators of $B(0|n)$ and 
the relations ${\cal R}$ consists of the triple para-Bose relations
given already in~(\ref{pBose}).

\smallskip\noindent {\bf Step 2.} 
If we delete two or more nodes from the distinguished Dynkin diagram, the resulting
$\Z$-gradings of $B(0|n)$ are no longer of the form~(\ref{5grading}).

\smallskip \noindent {\bf Step 3.}
Delete node $i$, $i=0,1,\ldots,n$ from the extended distinguished Dynkin diagram.
The remaining diagram is that of $G_0=B(0|n)$, $A_1\oplus B(0|n-1)$, 
$C_2\oplus B(0,n-2)$, $\ldots$,  $C_{n-1}\oplus B(0|1)$,
$C_n$.
In all these cases there is only one 
$G_0$-module, so there are no contributions to the classification.

\smallskip \noindent {\bf Step 4.} 
Delete the adjacent nodes $(i-1)$ and $i$, $i=2,3,\dots,n$ from the 
extended distinguished Dynkin diagram. The remaining diagram is that of 
$\tilde{G}_0=A_1\oplus B(0|n-2)$ for $i=2$, of
$\tilde{G}_0=C_{i-1}\oplus B(0|n-i)$ for $i=3,\ldots , n-1$ and of
$\tilde{G}_0=C_{n-1}$ for $i=n$. In each case there are 
seven $\tilde{G}_0$-modules $g_k$, one of which is invariant under $\omega$
(say $g_1$). Then one has to put $G_0=H+\tilde{G}_0+g_1$, and in each 
case one finds $G_0\equiv H+B(0|n-1)$. Now there are four $G_0$-modules and
$B(0|n)=G_{-2}\oplus G_{-1}\oplus G_0\oplus G_{+1} \oplus G_{+2}$ with 
\begin{eqnarray}
G_{-1}&=&\hbox{span}\{ e_{1, 1+n+i}+e_{1+i,1}, e_{1+i,1+n+j}+e_{1+j,1+n+i},\nn\\
&& e_{1+i, 1+j}-e_{n+1+j,n+1+i}, \ i\neq j =1,\ldots,n \}, \nn
\end{eqnarray}
$N=2n-1$, and $G_{-2}=\hbox{span}\{ e_{1+i,1+n+i}\}$.
All these cases are isomorphic to those of Step~1 with $i=1$.

\smallskip \noindent {\bf Step 5.} 
Delete the nonadjacent nodes $i,j$ from the extended distinguished Dynkin diagram.
The remaining diagram is that of 
$\tilde{G}_0=C_i\oplus sl(j-i)\oplus B(0|n-j)$ (for $i=1$,
$\tilde{G}_0=A_1\oplus sl(j-1)\oplus B(0|n-j)$; for $j=n$, 
$\tilde{G}_0=C_i\oplus sl(n-i)$).
In each case there are 
seven $\tilde{G}_0$-modules $g_k$, one of which is invariant under $\omega$
(say $g_1$). Then one has to put $G_0=H+\tilde{G}_0+g_1$, and in each 
case one finds $G_0\equiv H+sl(j-i)\oplus B(0|n-j+i)$. 
Now there are four $G_0$-modules and $B(0|n)=G_{-2}\oplus G_{-1}\oplus G_0\oplus G_{+1}\oplus G_{+2}$. 
All these cases are isomorphic to those in Step~1 with $i=2,\ldots,n-1$.

\smallskip \noindent {\bf Step 6.}
If we delete three or more nodes from the extended distinguished 
Dynkin diagram, the corresponding
$\Z$-grading of $B(0|n)$ has no longer the required properties (i.e.\ there are
non-zero subspaces $G_i$ with $|i|>2$).

In the case of $B(0|n)$ any other choice of simple roots is equivalent to the 
distinguished choice, so there are no more cases to study.

\section{The Lie superalgebras $D(m|n)$}
\setcounter{equation}{0} \label{sec:D}

$G=D(m|n)\equiv osp(2m|2n)$ is the subalgebra of $sl(2m|2n)$ consisting of matrices of the 
form~(\ref{osp}) with the middle row and column deleted.
The Cartan subalgebra $H$ of $G$ is again the
subspace of diagonal matrices $D$. 
Putting $\epsilon_i(D)=d_{ii}$, $i=1,\ldots, m$, 
$\delta_i(D)=d_{2m+i,2m+i}$, $i=1,\ldots,n$,
the even root vectors and corresponding roots of $G$ are given by:
\begin{eqnarray*}
e_{jk}-e_{k+m,j+m} & \leftrightarrow &\epsilon_j -\epsilon_k, \qquad j\neq k=1,\ldots ,m,\\
e_{j,k+m}-e_{k,j+m} & \leftrightarrow &\epsilon_j +\epsilon_k, \qquad j<k=1,\ldots ,m,\\
e_{j+m,k}-e_{k+m,j} & \leftrightarrow &-\epsilon_j -\epsilon_k, \qquad j<k=1,\ldots ,m,\\
e_{2m+j,2m+k}-e_{n+2m+k,n+2m+j} & \leftrightarrow &\delta_j -\delta_k, 
\qquad j\neq k=1,\ldots ,n,\\
e_{2m+j,2m+k+n}+e_{2m+k,2m+j+n} & \leftrightarrow 
&\delta_j +\delta_k, \qquad j\leq k=1,\ldots ,n,\\
e_{2m+n+j,2m+k}+e_{2m+n+k,2m+j} & \leftrightarrow &-
\delta_j -\delta_k, \qquad j\leq k=1,\ldots ,n,
\end{eqnarray*}
and the odd root vectors and roots by:
\begin{eqnarray*}
e_{j,2m+k}-e_{2m+n+k,j+m} & \leftrightarrow &\epsilon_j -\delta_k, 
\qquad j=1,\ldots ,m, \ k=1,\ldots ,n,\\
e_{m+j,2m+k}-e_{2m+n+k,j} & \leftrightarrow &-\epsilon_j -\delta_k, 
\qquad j=1,\ldots ,m, \ k=1,\ldots , n,\\
e_{j,2m+n+k}+e_{2m+k,m+j} & \leftrightarrow &\epsilon_j +\delta_k, 
\qquad j=1,\ldots ,m, \ k=1,\ldots, n,\\
e_{m+j,2m+n+k}+e_{2m+k,j} & \leftrightarrow &-\epsilon_j +\delta_k, 
\qquad j=1,\ldots ,m, \ k=1,\ldots, n.\\
\end{eqnarray*}

The distinguished set of simple roots, the corresponding Dynkin diagram and its extension
are given in Table~1.

\smallskip \noindent {\bf Step 1.} 
Delete node $i$, $i=1,\ldots,m+n-2$ from the distinguished Dynkin diagram. 
The corresponding diagram is that of $G_0=H+D(m|n-1)$ for $i=1$, of 
$G_0=H+sl(i)\oplus D(m|n-i)$ for $i=2,\ldots ,n-1$, of 
$G_0=H+sl(n)\oplus D_m$ for $i=n$, 
of $G_0=H+sl(i-n|n)\oplus D_{m+n-i}$ for $i=n+1, \ldots ,m+n-3$
and of $G_0=H+sl(m-2|n)\oplus A_1\oplus A_1$ for $i=m+n-2$.
There are four simple $G_0$-modules and $D(m|n)=G_{-2}\oplus G_{-1}\oplus G_0\oplus G_{+1}\oplus G_{+2}$,
where for, $i=1,\ldots n$:
\begin{eqnarray}
G_{-1}&=&\hbox{span}\{ e_{2m+j,2m+n+k}+e_{2m+k,2m+j+n}, e_{2m+j,2m+k}-e_{n+2m+k,n+2m+j},
\nn \\
&&\ e_{l,2m+n+j}+e_{2m+j,m+l},  e_{m+l,2m+n+j}+e_{2m+j,l};
\nn \\
&& j=1,\ldots ,i,\ k=i+1,\ldots ,n,\ l=1, \ldots, m\}, \label{D21}
\end{eqnarray} 
with $N=2i(m+n-i)$; whereas for $i=n+1, \ldots m+n-2$:
\begin{eqnarray}
G_{-1}&=&\hbox{span}\{ e_{k,2m+n+j}+e_{2m+j,m+k},  e_{m+k,2m+n+j}+e_{2m+j,k}, \nn\\
&& e_{l,m+k}-e_{k,m+l}, e_{lk}-e_{m+k,m+l};
\nn \\
&& \ j=1,\ldots ,n,\ k=i-n+1,\ldots ,m,\  l=1, \ldots, i-n,\}, \label{D22}
\end{eqnarray}
with $N=2i(m+n-i)$. 

\smallskip \noindent {\bf Step 2.} 
Delete node $m+n-1$ or $m+n$ from the distinguished Dynkin diagram. The corresponding diagram is 
the Dynkin diagram of $G_0=H+sl(m|n)$.
There are two simple $G_0$-modules and $D(m|n)= G_{-1}\oplus G_0\oplus G_{+1}$,
where, for $m+n-1$:
\begin{eqnarray}
G_{-1}&=&\hbox{span}\{ e_{j,m+k}-e_{k,m+j} ,\ j<k=1,\ldots,m-1,
\nn \\
&&\ e_{2m+q,2m+n+s}+e_{2m+s,2m+n+q}, \ q\leq s =1,\ldots ,n,
\nn \\
&& e_{lm}+e_{2m,m+l}, e_{l,2m+n+p}+e_{2m+p,m+l}, e_{2m,2m+n+p}+e_{2m+p,m}, \nn\\
&& l=1,\ldots,m-1, \  p=1,\ldots,n\}; \label{D22a}
\end{eqnarray}
and for $m+n$: 
\begin{eqnarray}
G_{-1}&=&\hbox{span}\{ e_{j,m+k}-e_{k,m+j},\ j<k=1,\ldots,m,
\nn \\
&&\ e_{2m+p,2m+n+s}+e_{2m+s,2m+n+p} , \ p\leq s =1,\ldots ,n,
\nn \\
&& e_{j,2m+n+l}+e_{2m+l,m+j}, \ j=1,\ldots,m, \ l=1,\ldots,n\}. \label{D22b}
\end{eqnarray}
Both cases have $N=(m+n)(m+n+1)/2-m$, and they are isomorphic.

\smallskip \noindent {\bf Step 3.}
Upon deleting two nodes $i$ and $j$ (except $i=m+n-1$, $j=m+n$) or more from the 
distinguished Dynkin diagram of $D(m|n)$, the corresponding $\Z$-gradings
have no longer the required property (there are non-zero $G_i$ with $|i|>2$).

\smallskip \noindent {\bf Step 4.} 
Delete node $m+n-1$ and $m+n$ from the distinguished Dynkin diagram.
We have $G_0=H+sl(m-1|n)$.
There are six simple $G_0$-modules. All the possible combinations 
of these modules give rise to gradings of the form~(\ref{5grading}).
There are essentially three different ways in which these $G_0$-modules
can be combined. To characterize these three cases, it is sufficient
to give only $G_{-1}$:
\begin{eqnarray}
G_{-1}&=&\hbox{span}\{ e_{jm}-e_{2m,m+j}, e_{j,2m}-e_{m,m+j}, e_{m,2m+k}-e_{2m+n+k,2m}, 
e_{m,2m+n+k}+e_{2m+k,2m};
\nn \\
&& \ j=1,\ldots ,m-1,\ k=1,\ldots ,n\}, \label{D41}
\end{eqnarray}
with $N=2(m+n-1)$;
\begin{eqnarray}
G_{-1}&=&\hbox{span}\{ e_{jm}-e_{2m,m+j}, e_{2m,2m+n+k}+e_{2m+k,m}, e_{m+j,l}-e_{m+l,j},
\nn \\
&& e_{m+j,2m+k}-e_{2m+n+k,j}, e_{2m+n+k,2m+p}+e_{2m+n+p,2m+k};  \ j=1,\ldots, m-1,\nn\\
&& k=1,\ldots, n,\ j<l=1,\ldots,m-1,\ k\leq p=1,\ldots,n\}, \label{D42}
\end{eqnarray}
with $N=(m+n)(m+n+1)/2-m$; 
\begin{eqnarray}
G_{-1}&=&\hbox{span}\{ e_{j,m+l}-e_{l,m+j}, e_{j,2m+n+k}+e_{2m+k,m+j}, e_{2m+k,2m+n+p}+e_{2m+p,2m+n+k},\nn\\
&& e_{m+j,m}-e_{2m,j} , e_{2m,2m+k}-e_{2m+n+k,m}; \ j=1,\ldots, m-1,\nn\\
&& k=1,\ldots,  n,\ j<l=1,\ldots,m-1,\ k\leq p=1,\ldots,n\}, \label{D43}
\end{eqnarray}
with $N=(m+n)(m+n+1)/2-m$. 
The cases (\ref{D42}) and (\ref{D43}) are isomorphic.

\smallskip \noindent {\bf Step 5.}
Delete node $i$, $i=0,1,\ldots,n+m$ from the extended distinguished Dynkin diagram.
The remaining diagram is that of $G_0=D(m|n)$, $A_1\oplus D(m|n-1)$,
$C_2\oplus D(m|n-2)$, $\ldots$, $C_{n-1}\oplus D(m|1)$,
$C_n\oplus D_m$, $C(n+1)\oplus D_{m-1}$, $D(2|n)\oplus D_{m-2}$, $\ldots$, 
$D(m-2|n)\oplus A_1\oplus A_1$, $D(m|n)$, $D(m|n)$.
In all these cases there is only one 
$G_0$-module, so there are no contributions to the classification.

\smallskip \noindent {\bf Step 6.} 
Delete the adjacent nodes $(i-1)$ and $i$, $i=2,3,\dots,n$ from the 
extended distinguished Dynkin diagram. The remaining diagram is that of 
$\tilde{G}_0=A_1\oplus D(m|n-2)$ for $i=2$, of
$\tilde{G}_0=C_{i-1}\oplus D(m|n-i)$ for $i=3,\ldots,n-1$ and of
$\tilde{G}_0=C_{n-1}\oplus D_m$ for $i=n$. In each case there are 
seven $\tilde{G}_0$-modules $g_k$, one of which is invariant under $\omega$
(say $g_1$). Then one has to put $G_0=H+\tilde{G}_0+g_1$, and in each 
case one finds $G_0\equiv H+D(m|n-1)$. Now there are four $G_0$-modules and
$D(m|n)=G_{-2}\oplus G_{-1}\oplus G_0\oplus G_{+1} \oplus G_{+2}$ with 
\begin{eqnarray}
G_{-1}&=&\hbox{span}\{ e_{2m+k, 2m+n+i}+e_{2m+i,2m+n+k}, e_{2m+i,2m+k}-e_{2m+n+k,2m+n+i},
\ i\neq k =1,\ldots,n,\nn\\
&& e_{l, 2m+n+i}+e_{2m+i,m+l}, e_{m+l, 2m+n+i}+e_{2m+i,l}, 
\ l=1,\ldots,m\}, \nn
\end{eqnarray}
$N=2(m+n-1)$ and $G_{-2}=\hbox{span}\{ e_{2m+i,2m+n+i}\}$.
All these cases are isomorphic to that of Step~1 with $i=1$.

\smallskip \noindent {\bf Step 7.} 
Delete the adjacent nodes $(i-1)$ and $i$, $i=n+1,\dots,m+n-1$ from the 
extended distinguished Dynkin diagram. The remaining diagram is that of 
$\tilde{G}_0=C_{n}\oplus D_{m-1}$ for $i=n+1$, of
$\tilde{G}_0=C(n+1)\oplus D_{m-2}$ for $i=n+2$, of
$\tilde{G}_0=D(i-n-1|n)\oplus D_{m+n-i}$ for $i=n+3,\ldots, m+n-3$,
of $\tilde{G}_0=D(m-3|n)\oplus A_1\oplus A_1$ for $i=m+n-2$
and of $\tilde{G}_0=D(m-2|n)\oplus A_1$ for $i=m+n-1$. In each case there are 
five $\tilde{G}_0$-modules $g_k$, one of which is invariant under $\omega$
(say $g_1$). Then one has to put $G_0=H+\tilde{G}_0+g_1$, and in each 
case one finds $G_0\equiv H+D(m-1|n)$. Now there are two $G_0$-modules and
$D(m|n)=G_{-1}\oplus G_0\oplus G_{+1}$ with 
\begin{eqnarray}
G_{-1}&=&\hbox{span}\{ e_{i-n, 2m+n+j}+e_{2m+j,m+i-n}, e_{i-n,2m+j}-e_{2m+n+j,m+i-n},
\ j =1,\ldots,n,\nn\\
&& e_{k, m+i-n}-e_{i-n,m+k}, e_{i-n,k}-e_{m+k,m+i-n}, 
\ i-n\neq k=1,\ldots,m \} \label{D7}
\end{eqnarray}
and $N=2(m+n-1)$. All these cases are mutually isomorphic.
 
\smallskip \noindent {\bf Step 8.} 
Delete the adjacent nodes $m+n-1$ and $m+n$ from the 
extended distinguished Dynkin diagram. The remaining diagram is that of 
$G_0\equiv D(m-1|n)$. There are two $G_0$-modules and
$D(m|n)=G_{-1}\oplus G_0\oplus G_{+1}$ with 
\begin{eqnarray}
G_{-1}&=&\hbox{span}\{ e_{m, 2m+n+j}+e_{2m+j,2m}, e_{m,2m+j}-e_{2m+n+j,2m},
\ j =1,\ldots,n,\nn\\
&& e_{k, 2m}-e_{m,m+k}, e_{m,k}-e_{m+k,2m}, 
\ k=1,\ldots,m-1,\}, \label{D8}
\end{eqnarray}
with $N=2(m+n-1)$. 
This case and all cases from Step~7 are isomorphic.

\smallskip \noindent {\bf Step 9.} 
Delete the nonadjacent nodes $i,j$, $i<j-1$, $j=3,\ldots,n$
from the extended distinguished Dynkin diagram.
The remaining diagram is that of 
$\tilde{G}_0=C_i\oplus sl(j-i)\oplus D(m|n-j)$ (for $i=1$
instead of $C_i$ we have $A_1$).
In each case there are 
seven $\tilde{G}_0$-modules $g_k$, one of which is invariant under $\omega$
(say $g_1$). Then one has to put $G_0=H+\tilde{G}_0+g_1$, and in each 
case one finds $G_0\equiv H+sl(j-i)\oplus D(m|n-j+i)$. 
Now there are four $G_0$-modules and $D(m|n)=G_{-2}\oplus G_{-1}\oplus G_0\oplus G_{+1}\oplus G_{+2}$. 
All these cases are isomorphic to those of Step~1 with $i=2,\ldots,n-1$.

\smallskip \noindent {\bf Step 10.} 
Delete the nonadjacent nodes $i,j$, $i=1,\ldots,n$,  $j=n+1,\ldots,n+m-2$
[but $(i,j)\ne(n,n+1)$] from the extended distinguished Dynkin diagram.
The remaining diagram is that of 
$\tilde{G}_0=C_i\oplus sl(j-n|n-i)\oplus D_{m+n-j}$ (for $j=n+m-2$
instead of $D_{n+m-j}$ we have $A_1\oplus A_1$).
In each case there are 
seven $\tilde{G}_0$-modules $g_k$, one of which is invariant under $\omega$
(say $g_1$). Then one has to put $G_0=H+\tilde{G}_0+g_1$, and in each 
case one finds $G_0\equiv H+sl(j-n|n-i)\oplus D(m+n-j|i)$. 
Now there are four $G_0$-modules and $D(m|n)=G_{-2}\oplus G_{-1}\oplus G_0\oplus G_{+1}\oplus G_{+2}$
with
\begin{eqnarray}
G_{-1}&=&\hbox{span}\{ e_{kl}-e_{m+l,m+k}, e_{k,m+l}-e_{l,m+k}, e_{r, 2m+n+s}+e_{2m+s,m+r}, \nn\\
&& e_{m+l,2m+n+p}+e_{2m+p,l}, e_{k,2m+q}-e_{2m+n+q,m+k}, \nn\\
&& e_{2m+q,2m+n+p}+e_{2m+p,2m+n+q}, e_{2m+p,2m+q}-e_{2m+n+q,2m+n+p},\nn\\
&&\ k=1,\ldots,j-n,\ l=j-n+1,\ldots,m, \ r=1,\ldots,m,\ s=1,\ldots, n, \nn\\
&&\ q=1,\ldots,i,\ p=i+1,\ldots,n\} ,  \label{D10}
\end{eqnarray}
with $N=2(j-i)(m+n-j+i)$.

\smallskip \noindent {\bf Step 11.} 
Delete the nonadjacent nodes $i,j$, $i=1,\ldots,n$, and $j=n+m-1$ or $n+m$
from the extended distinguished Dynkin diagram.
The remaining diagram is that of 
$G_0=C_i\oplus sl(m|n-i)$. 
There are four $G_0$-modules and $D(m|n)=G_{-2}\oplus G_{-1}\oplus G_0\oplus G_{+1}\oplus G_{+2}$
with
\begin{eqnarray}
G_{-1}&=&\hbox{span}\{ e_{k,2m+p}-e_{2m+n+p,m+k}, e_{k,2m+n+p}+e_{2m+p,m+k}, \nn\\
&& e_{2m+p,2m+n+q}+e_{2m+q,2m+n+p}, e_{2m+q,2m+p}-e_{2m+n+p,2m+n+q}; \nn\\
&& \ k=1,\ldots,m,\ p=1,\ldots,i, \ q=i+1,\ldots,n\} \label{D11}
\end{eqnarray}
for $j=m+n$ and a similar expression for $j=m+n-1$. Naturally, both cases
are isomorphic and $N=2i(m+n-i)$.

\smallskip \noindent {\bf Step 12.} 
Delete the nonadjacent nodes $i,j$, $i=n+1,\ldots,m+n-2$, $j=n+m-1$ or $n+m$
from the extended distinguished Dynkin diagram.
The remaining diagram is that of 
$G_0=D(i-n|n)\oplus sl(m+n-i)$ (if $i=n+1$ instead of $D(i-n|n)$ we have $C(n+1)$). 
There are four $G_0$-modules and $D(m|n)=G_{-2}\oplus G_{-1}\oplus G_0\oplus G_{+1}\oplus G_{+2}$
with
\begin{eqnarray}
G_{-1}&=&\hbox{span}\{ e_{kl}-e_{m+l,m+k}, e_{m+k,l}-e_{m+l,k}, e_{m+l,2m+n+p}+e_{2m+p,l}; 
e_{m+l,2m+p}-e_{2m+n+p,l}, \nn\\
&&\ k=1,\ldots,i-n,\ l=i-n+1,\ldots,m, p=1,\ldots,n \}, \label{D12}
\end{eqnarray}
for $j=m+n$ and a similar expression for $j=m+n-1$, both having $N=2i(m+n-i)$. 
The cases here and in Step~10 with one and the same $G_0$ and $N$ are 
isomorphic. 

\smallskip \noindent {\bf Step 13.} 
Delete the nonadjacent nodes $i,j$, $i<j-1$, $i=n+1,\ldots,n+m-4$,  
$j=n+3,\ldots,n+m-2$ from the extended distinguished Dynkin diagram.
The remaining diagram is that of 
$\tilde{G}_0=D(i-n|n)\oplus sl(j-i)\oplus D_{m+n-j}$ (for $i=n+1$
instead of $D(i-n|n)$ we have $C(n+1)$, for $j=m+n-2$ instead of 
$D_{m+n-j}$ we have $A_1\oplus A_1$).
In each case there are 
seven $\tilde{G}_0$-modules $g_k$, one of which is invariant under $\omega$
(say $g_1$). Then one has to put $G_0=H+\tilde{G}_0+g_1$, and in each 
case one finds $G_0\equiv H+sl(j-i)\oplus D(m-j+i|n)$. 
Now there are four $G_0$-modules and $D(m|n)=G_{-2}\oplus G_{-1}\oplus G_0\oplus G_{+1}\oplus G_{+2}$
with
\begin{eqnarray}
G_{-1}&=&\hbox{span}\{ e_{kl}-e_{m+l,m+k}, e_{m+k,l}-e_{m+l,k}, e_{m+l, 2m+n+q}+e_{2m+q,l}, 
e_{m+l,2m+q}-e_{2m+n+q,l}, \nn\\
&&\ k=1,\ldots,i-n,j-n+1,\ldots ,m, \ l=i-n+1,\ldots,j-n, q=1,\ldots,n\}  \nn
\end{eqnarray}
with $N=2(j-i)(m+n-j+i)$. All of these are isomorphic to cases in Step~12.

\smallskip \noindent {\bf Step 14.}
If we delete three or more nodes from the 
extended distinguished Dynkin diagram, the corresponding
$\Z$-grading of $D(m|n)$ has no longer the required properties (i.e.\ there are
non-zero subspaces $G_i$ with $|i|>2$).

\smallskip \noindent {\bf Step 15.}
Also here, we have considered all the non-distinguished Dynkin
diagrams and their extensions, following~\cite{Serganova}.
Repeating our procedure for all of them, leads again to a lot of case studies,
most of which are isomorphic to results of the previous steps.
There is, however, one extra case that is not covered in the previous steps, and that we
shall briefly describe.
Consider the non-distinguished Dynkin diagrams of $D(m|n)$ of the following form:
\[
\includegraphics{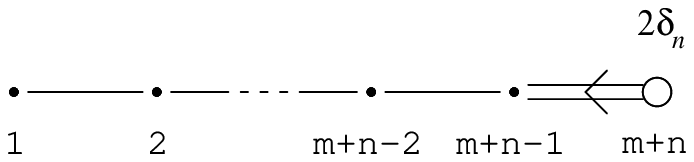}
\]
In this diagram, each dot stands for a white or gray circle, depending upon whether the corresponding
simple root is even or odd. All these other simple roots are of the form $\eta-\eta'$, where $\eta$ and $\eta'$
can be a $\epsilon_i$ or a $\delta_j$. For example, root $m+n-1$ is either $\delta_{n-1}-\delta_n$ (in which
case the circle is white) or $\epsilon_m-\delta_n$ (in which case it is gray).
Deleting node $i$ ($i=1,\ldots,m+n-2$) yields the (non-distinguished) Dynkin diagram of $G_0=sl(k|l)\oplus
D(m-k|n-l)$. Most of these cases are isomorphic to those already found in Step~10 or~11. However, 
the case $G_0=H+sl(m-1|l)\oplus D(1|n-l)=H+sl(m-1|l)\oplus C(n-l+1)$ ($l=0,\ldots,n-1$) did not occur before. 
This adds a new case to the classification, for which the length of the grading is~5.
{}From a detailed analysis, it follows that this is the only extra case that can be obtained from 
the non-distinguished Dynkin diagrams (extended or not).

\section{The Lie superalgebras $C(n)$}
\setcounter{equation}{0} \label{sec:C}

Let $G=C(n)=D(1|n-1)=osp(2|2n-2)$. For a description of the
root vectors, we refer to the previous section. The even roots are
of the form $\pm \delta_j\pm\delta_k$ ($j,k=1,\ldots,n-1$) and the
odd roots are $\pm\epsilon \pm\delta_k$ ($k=1,\ldots,n-1$).
This Lie superalgebra is treated separately from $D(m|n)$ because its distinguished
Dynkin diagram is different (see Table~1), and its structure is also different
(it is a type~I Lie superalgebra~\cite{Kac}).

\smallskip \noindent {\bf Step 1.} 
Delete node $1$ from the distinguished 
Dynkin diagram. Then the corresponding diagram is 
that of $G_0=C_{n-1}$. There are only two $G_0$-modules and 
$C(n)$ has the grading $C(n)=G_{-1}\oplus G_0\oplus G_{+1}$,
where
\begin{eqnarray}
G_{-1}&=&\hbox{span}\{ e_{1,i+2}-e_{n+1+i,2}, e_{1,n+1+i}+e_{i+2,2}, \ i=1,\ldots,n-1\},\label{C1} \\
G_{+1}&=&\hbox{span}\{ e_{2,n+1+i}+e_{i+2,1}, e_{2,i+2}-e_{n+1+i,1},\ i=1,\ldots,n-1\}, \nn
\end{eqnarray}
with $N=2(n-1)$. Putting 
\begin{eqnarray}
&& c_{-,i}^-=e_{1,2+i}-e_{n+1+i,2}, \qquad c_{+,i}^-=e_{1,n+1+i}+e_{2+i,2}, \nn\\
&& c_{-,i}^+=e_{2,n+1+i}+e_{2+i,1}, \qquad c_{+,i}^+=e_{2,2+i}-e_{n+1+i,1}, \nn
\end{eqnarray}
the operators $c_{\xi i}^\pm$, $\xi =\pm$, $i=1,\ldots , n-1$,
satisfy the following relations:
\begin{eqnarray}
&&[\{ c_{\xi i}^-, c_{\eta j}^+\} , c_{\epsilon k}^+]=\xi \delta_{\xi \eta}
\delta_{ij}c_{\epsilon k}^+ -\epsilon \delta_{\xi \epsilon}\delta_{ik}c_{\eta j}^+ 
+\eta \delta_{-\eta \epsilon}\delta_{jk}c_{-\xi i}^+ ,\nn\\
&& [\{ c_{\xi i}^-, c_{\eta j}^+\} , c_{\epsilon k}^-]=-\xi \delta_{\xi \eta}
\delta_{ij}c_{\epsilon, k}^- +\eta \delta_{\eta \epsilon}\delta_{jk}c_{\xi i}^- 
+\epsilon \delta_{-\xi \epsilon}\delta_{ik}c_{-\eta j}^- ,\nn\\
&&\{ c_{-i}^-, c_{+j}^+\}=\{ c_{-j}^-, c_{+i}^+\},\qquad 
\{ c_{+i}^-, c_{-j}^+\}=\{ c_{+j}^-, c_{-i}^+\},\nn\\
&& \{ c_{\xi i}^-, c_{\eta j}^-\}=\{ c_{\xi i}^+, c_{\eta j}^+\}=0 .\label{CS1}
\end{eqnarray}
(here and throughout, by convention, $\xi, \eta, \epsilon$ are written as $\pm $
when used as subscripts, and as $\pm 1$ when used algebraically as multipliers).

\smallskip \noindent {\bf Step 2.} 
Delete node $i$, $i=2,\ldots , n-1$, from the distinguished Dynkin diagram. 
The corresponding diagram is 
the Dynkin diagram of $G_0=H+sl(1|i-1)\oplus C_{n-i}$ (if $i=n-1$, $C_{n-i}=A_1$).
There are four simple $G_0$-modules and $C(n)=G_{-2}\oplus G_{-1}\oplus G_0\oplus G_{+1}\oplus G_{+2}$
where 
\begin{eqnarray}
G_{-1}&=&\hbox{span}\{ e_{j+2,k+2}-e_{n+1+k,n+1+j}, e_{j+2,n+1+k}+e_{k+2,n+1+j}, 
e_{1,k+2}-e_{n+1+k,2}, \nn\\
&& e_{1,n+1+k}+e_{k+2,2},;  \ j=1,\ldots,i-1,\ k=i,\ldots,n-1 \},\label{C2}
\end{eqnarray}
with $N=2i(n-i)$. An interesting case is that with $i=n-1$ and $N=2(n-1)$.

\smallskip \noindent {\bf Step 3.} Delete node $n$ from the distinguished 
Dynkin diagram. Then the corresponding diagram is 
that of $G_0=sl(1|n-1)$. There are only two $G_0$-modules and 
$C(n)$ has grading $C(n)=G_{-1}\oplus G_0\oplus G_{+1}$, where
\begin{eqnarray}
G_{-1}&=&\hbox{span}\{ e_{k+2,n+1+l}+e_{l+2,n+1+k}, \ k\leq l=1,\ldots,n-1, \nn\\
&& e_{1,n+1+j}+e_{j+2,2}, \ j=1,\ldots,n-1 \}.\label{C3}
\end{eqnarray}
This is also an interesting case, since there are $N=\frac{n(n+1)}{2}-1$ supercommuting 
annihilation (resp.\ creation) operators.

\smallskip \noindent {\bf Step 4.}
Upon deleting two nodes $i$ and $j$ (except $i=1$, $j=n$) 
or more from the distinguished Dynkin diagram of $C(n)$, the corresponding $\Z$-gradings
have no longer the required property (there are non-zero $G_i$ with $|i|>2$).

\smallskip \noindent {\bf Step 5.} 
Delete node $1$ and $n$ from the distinguished Dynkin diagram.
We have $G_0=H+sl(n-1)$.
There are six simple $G_0$-modules. All the possible combinations 
of these modules give rise to gradings of the form
$C(n)=G_{-2}\oplus G_{-1}\oplus G_0\oplus G_{+1}\oplus G_{+2}$.
There are essentially three different ways in which these $G_0$-modules
can be combined. To characterize these three cases, it is sufficient
to give only $G_{-1}$:
\begin{eqnarray}
G_{-1}&=&\hbox{span}\{ e_{j+2,n+k+1}+e_{k+2,n+j+1},\ j\leq k =1,\ldots , n-1, \nn \\
&& e_{1,i+2}-e_{n+i+1,2},\ i=1,\ldots,n-1\} ,\label{c5.1} 
\end{eqnarray}
with $N=n(n+1)/2-1$;
\begin{eqnarray}
G_{-1}&=&\hbox{span}\{ e_{1,k+2}-e_{n+k+1,2}, e_{2,k+2}-e_{n+k+1,1}, k=1,\ldots,n-1\} ,
\label{c5.2} 
\end{eqnarray}
with $N=2(n-1)$; 
\begin{eqnarray}
G_{-1}&=&\hbox{span}\{ e_{1,n+k+1}+e_{k+2,2},\ k =1,\ldots , n-1, \nn \\
&& e_{n+l+1,p+2}+e_{n+p+1,l+2},\ l\leq p=1,\ldots,n-1\} ,\label{c5.3} 
\end{eqnarray}
with $N=n(n+1)/2-1$. 
It is interesting to give the algebraic relations for~(\ref{c5.2}), since the number of creation and annihilation operators is $N=2(n-1)$. One can label the CAOs as follows ($k=1,\ldots,n-1$):
\begin{eqnarray}
&& c_{-k}^-=e_{1,k+2}-e_{n+k+1,2}, \qquad c_{+k}^-=e_{2,k+2} -e_{n+k+1,1},\nn \\
&& c_{-k}^+=e_{2,n+k+1}+e_{k+2,1}, \qquad c_{+k}^+=e_{1,n+k+1}+e_{k+2,2}. \nn
\end{eqnarray}
The CAOs $c_{\xi k}^\pm$, $\xi =\pm$, $k=1,\ldots, n-1$,
satisfy the relations ($\xi, \eta, \epsilon =\pm$  or $\pm 1$; $j,k,l=1,\ldots,n-1$):
\begin{eqnarray}
&&\{ c_{\xi j}^\eta ,c_{\xi k}^\eta\}=\{c_{- j}^-,c_{+ k}^+\}=
\{ c_{+j}^-, c_{-k}^+\}=0,\nn\\
&& \{c_{+ j}^-,c_{+ k}^+\}=\{c_{-j}^-, c_{-k}^+\}, \qquad j\neq k,  \nn \\
&& \{c_{\xi j}^\xi,c_{-\xi k}^\xi\}=\{c_{\xi k}^\xi,c_{-\xi j}^\xi\}, \qquad  \nn \\
&&[\{ c_{\xi j}^\gamma,c_{\eta k}^\gamma\},c_{\epsilon l}^\gamma ]=0, \nn \\
&&[\{ c_{\xi j}^\xi,c_{-\xi k}^\xi\},c_{\epsilon l}^{-\xi}]=
-\xi \delta_{kl}
 c_{-\epsilon j}^\xi -\xi \delta_{jl}c_{-\epsilon k}^\xi, \nn \\
&&[\{ c_{\xi j}^-,c_{\xi k}^+\},c_{\eta l}^-]=
-\delta_{kl}
 c_{\eta j}^- -(-1)^{\delta_{\xi \eta}} \delta_{jk}c_{\eta l}^-, \nn\\
 &&[\{ c_{\xi j}^-,c_{\xi k}^+\},c_{\eta l}^+ ]=
(-1)^{\delta_{\xi \eta}} \delta_{jk}
 c_{\eta l}^+ + \delta_{jl}c_{\eta k}^+. \label{CS2}
\end{eqnarray}
 
\smallskip \noindent {\bf Step 6.}
Delete node $i$, $i=0,1,\ldots,n$, from the extended distinguished Dynkin diagram.
The remaining diagram is that of $G_0=C(n)$, $C(n)$, $sl(2|1) \oplus C_{n-2}$,
$C(3)\oplus C_{n-3}$, $\ldots$, $C_{n-1}\oplus A_1$, $C(n)$.
In all these cases there is only one 
$G_0$-module, so there are no contributions to the classification.

\smallskip \noindent {\bf Step 7.} 
Delete the adjacent nodes $i$ and $i+1$, $i=2,3,\dots,n-2$, from the 
extended distinguished Dynkin diagram. The remaining diagram is that of 
$\tilde{G}_0=H+sl(2|1)\oplus C_{n-3}$ for $i=2$ and of
$\tilde{G}_0=H+C(i)\oplus C_{n-i-1}$ for $i=3,\ldots , n-2$. In each case there are 
seven $\tilde{G}_0$-modules $g_k$, one of which is invariant under $\omega$
(say $g_1$). Then one has to put $G_0=H+\tilde{G}_0+g_1$, and in each 
case one finds $G_0\equiv H+C(n-1)$. Now there are four $G_0$-modules and
$C(n)=G_{-2}\oplus G_{-1}\oplus G_0\oplus G_{+1} \oplus G_{+2}$ with 
\begin{eqnarray}
G_{-1}&=&\hbox{span}\{ e_{i+2, n+k+1}+e_{k+2,n+i+1}, e_{i+2,k+2}-e_{n+k+1,n+i+1}, 
e_{1, n+i+1}+e_{i+2,2},\nn\\
&& e_{2, n+i+1}+e_{i+2,1}; \ k\neq i=1,\ldots,n-1\} \label{C7}
\end{eqnarray}
and $N=2(n-1)$. All these cases are mutually isomorphic.
 
\smallskip \noindent {\bf Step 8.} 
Delete the adjacent nodes $n-1$ and $n$ from the 
extended distinguished Dynkin diagram. The remaining diagram is that of 
$G_0\equiv H+C(n-1)$. This case turns out to be isomorphic to those of Step~7.

\smallskip \noindent {\bf Step 9.} 
Delete the nonadjacent nodes $i$, $i=2,3,\dots,n-2$, and 
$j=n$ from the extended distinguished Dynkin diagram. The remaining diagram is that of 
$G_0=H+C(i)\oplus sl(n-i)$ (for $i=2$ 
instead of $C(i)$ we have $sl(2|1)$).
In each case there are four $G_0$-modules and 
$C(n)=G_{-2}\oplus G_{-1}\oplus G_0\oplus G_{+1} \oplus G_{+2}$ with 
\begin{eqnarray}
G_{-1}&=&\hbox{span}\{ e_{k+2,l+2}-e_{n+1+l,n+1+k}, e_{n+1+k,l+2}+e_{n+l+1,k+2}, 
e_{1, l+2}-e_{n+l+1,2}, \nn\\
&& e_{2, l+2}-e_{n+l+1,1};  l=i,\ldots,n-1,\ k=1,\ldots,i-1 \} \label{C9}
\end{eqnarray}
and $N=2i(n-i)$. These are all new cases.

\smallskip \noindent {\bf Step 10.} 
Delete the nonadjacent nodes $i<j-1$, $i=2,3,\dots,n-2$, 
$j=4,\ldots, n-1$ from the 
extended distinguished Dynkin diagram. The remaining diagram is that of 
$\tilde{G}_0=H+C(i)\oplus sl(j-i)\oplus C_{n-j}$ (for $i=2$ 
instead of $C(i)$ we have $sl(2|1)$, for $j=n-1$ instead of $C_{n-j}$
we have $A_1$).
In each case there are 
seven $\tilde{G}_0$-modules $g_k$, one of which is invariant under $\omega$
(say $g_1$). Then one has to put $G_0=H+\tilde{G}_0+g_1$, and in each 
case one finds $G_0\equiv H+sl(j-i)\oplus C(n-j+i)$. Now there are four $G_0$-modules and
$C(n)=G_{-2}\oplus G_{-1}\oplus G_0\oplus G_{+1} \oplus G_{+2}$ with 
\begin{eqnarray}
G_{-1}&=&\hbox{span}\{ e_{k+2,l+2}-e_{n+1+l,n+1+k}, e_{n+1+k,l+2}+e_{n+l+1,k+2}, 
e_{1, l+2}-e_{n+l+1,2}, \nn\\
&& e_{2, l+2}-e_{n+l+1,1}; l=i,\ldots,j-1,\ k=1,\ldots,i-1,j,\ldots, n-1 \} \label{C10}
\end{eqnarray}
and $N=2(j-i)(n-j+i)$. All these cases are amongst those of Step~10.

\smallskip \noindent {\bf Step 11.}
Upon deleting three or more  nodes from the 
extended distinguished Dynkin diagram of $C(n)$, the corresponding $\Z$-gradings
have no longer the required property (there are non-zero $G_i$ with $|i|>2$).  

The other non-distinguished choices for the simple root systems give no new results.

\setcounter{equation}{0}
\section{Summary and conclusions} \label{sec:concl}%

Our analysis has led to a complete classification of all GQS associated with 
the basic classical Lie superalgebras. Some cases in our classification
have appeared earlier as examples, e.g.\ para-Bose statistics~(\ref{B021}), 
$A$-superstatistics~(\ref{A1}) in~\cite{Palev5}, \cite{sl(1|n)}, 
and the combined para-Bose/para-Fermi case~(\ref{GB3})
in~\cite{Posp}.
Some other examples are also rather simple, e.g.\ the alternatives to
$A$-superstatistics in~(\ref{A2}), a statistics with two kind of
particles in~(\ref{Adouble}), and the statistics related to $C(n)$ 
superalgebras~(\ref{CS1}) and~(\ref{CS2}).

Although the detailed analysis in the previous sections was necessary to present
the complete solution, it is convenient to summarize the final results in 
a table. Table~2 recapitulates the classification of all GQS. From this table,
it follows that many of the earlier cases can be somewhat unified in a simple form,
provided one makes use of the common isomorphisms for Lie algebras and Lie superalgebras
(such as $D_1=C_1=A_1$, $D_2=A_1\oplus A_1$, $B(m|0)=B_m$, $D(1|1)=C(2)=sl(1|2)$ etc.).
For $A(m|n)$, there are essentially two distinct cases. Either $G_0=H+sl(k|l)\oplus sl(p|q)$,
in which case the grading has length~3 and $G_{\pm 1}$ is fixed by $G_0$. 
Or else $G_0=H+sl(k|l)\oplus sl(p|q)\oplus sl(r|s)$, in which case the grading has
length~5. In this second case, there are always three ways of combining the $G_0$-modules
in order to give some $G_{-1}$.
For $B(m|n)$, all the cases are characterized by a $G_0$ of the form
$G_0=H+sl(k|l)\oplus B(m-k,n-l)$. This includes cases such as $sl(l)\oplus B(m|n-l)$,
$sl(k|n)\oplus B_{m-k}$ (Step~1), $B(m|n-1)$ (Step~4) and $B(m-1|n)$ (Step~5).
Also the results for the remaining Lie superalgebras can be neatly summarized.
Note that for $D(m|n)$ (and for $C(n)=D(1|n-1)$) there is one $G_0$ which gives rise
not only to different possibilities for $G_{-1}$ but even for $N$.

A striking property, see Table~2, is that all basic classical Lie superalgebras, except $B(0|n)$, 
allow a GQS with a grading of length~3; in other words, a GQS with supercommuting creation
and annihilation operators. 

Note that a set of CAOs together with a complete set
of relations ${\cal R}$ unambiguously describes the Lie superalgebra. So each case
of our classification also gives the description of a Lie superalgebra
in terms of a number of generators subject to certain relations.
This can also be reformulated in terms of the notion of Lie supertriple 
systems~\cite{Okubo}. In fact, in our case the subspace $G_{-1}\oplus G_{+1}$
(i.e.\ the subspace spanned by all CAOs) is a Lie supertriple system for
the universal enveloping algebra $U(G)$.

Just as in~\cite{GQS}, we have dealt only with a mathematical definition
of generalized quantum statistics. 
In order to talk about a quantum statistics in the physical sense,
one should take into account additional requirements for the
CAOs, related to certain quantization postulates~\cite{Palev2}.
These conditions are related to the existence of state spaces (Fock spaces),
in which the CAOs act in such a way that the corresponding observables
are Hermitian operators.
We refer to section~VII of~\cite{GQS} for a discussion on this.
We hope that some cases of our classification will yield interesting
GQS also from this point of view.

As a second application, we mention the possible solutions of Wigner Quantum Systems~\cite{WQS1}.
Roughly speaking, the compatibility conditions (CCs) to be satisfied by a Wigner Quantum Oscillator
system (see formula (3.7) in~\cite{WQS2}) are written in terms of certain odd operators $A_i^\pm$; furthermore,
these CCs are special triple relations. 
So it is of importance to investigate which triple relations ${\cal R}$ of our
current classification of GQS could provide special solutions of these CCs.
It is known, for instance, that there is a $sl(1|n)$ solution~\cite{WQS1}-\cite{WQS3} 
or a $sl(n|3)$ solution~\cite{WQS4}.
Obviously, for a possible candidate solution all the CAOs of ${\cal R}$ should be odd
operators. Let us briefly describe the GQSs of the current classification which have
only odd CAOs. Then $G_{-1}$ and $G_{+1}$ are odd subspaces, and by the grading condition it
follows automatically that the GQS grading~(\ref{5grading}) is {\em consistent} with the
$\Z_2$-grading. So our problem reduces to selecting those GQS from our classification with
a consistent $\Z$-grading. This is not too difficult.

For $A(m|n)$, one can consider $i=m+1$ in Step~1. Then all elements of $G_{-1}$ in~(\ref{exA1})
are odd; $G_0=H+sl(m+1)\oplus sl(n+1)$, and the grading has length~3 (the case $m=0$ corresponds
to~\cite{Palev5}, \cite{WQS1}-\cite{WQS3}). Alternatively, one can consider the cases $(i,j)=(i,m+1)$ or $(i,j)=(m+1,j)$ ($i<j$)
of Step~2. In the case $(i,j)=(i,m+1)$, the elements of $G_{-1}$ given by~(\ref{A23}) are all odd,
and $G_0=H+sl(i)\oplus sl(m+1-i)\oplus sl(n+1)$. In the case $(i,j)=(m+1,j)$, 
the elements of $G_{-1}$ given by~(\ref{A22}) are all odd,
and $G_0=H+sl(m+1)\oplus sl(k)\oplus sl(n+1-k)$ with $k=j-m-1$. 
For $B(m|n)$, the case $i=n$ in Step~1 leads to a $G_{-1}$ with only odd elements in~(\ref{B21}).
Note that in this case $G_0=H+sl(n)\oplus B_m$.
For $B(0|n)$, this corresponds to taking $i=n$ in Step~1, so that $G_{-1}$ in~(\ref{B021}) has
odd elements only and $G_0=H+sl(n)$ (this is the para-Bose case).
For $D(m|n)$, the case $i=n$ in Step~1 leads to a $G_{-1}$ in~(\ref{D21}) with odd elements only, 
and with $G_0=H+sl(n)\oplus D_m$. There is a second solution here, namely the case $i=n$ in Step~11;
then $G_{-1}$, given by~(\ref{D11}), has odd elements only, and $G_0=H+sl(m)\oplus C_n$.
Finally, for $C(n)$ the solution provided in Step~1 has only odd elements for $G_{-1}$,
see~(\ref{C1}), with $G_0=H+C_{n-1}$. Also here there is a second solution, given in Step~5
with $G_{-1}$ of the form~(\ref{c5.2}) and $G_0=H+sl(n-1)$.
To conclude, all the basic classical Lie superalgebras have GQSs with odd CAOs only. 
Whether all these cases provide special solutions to the CCs of the Wigner Quantum Oscillator
will be treated elsewhere.

\bigskip
\noindent{\bf Acknowledgments}
\medskip

\noindent

N.I.\ Stoilova was supported by a project from the Fund for Scientific Research -- Flanders (Belgium).

\newpage
\noindent
{\bf Table 1}.
Classical Lie superalgebras, their (extended) Dynkin diagrams with a labeling of
the nodes and the corresponding simple roots.
\vskip 1mm
\noindent
\begin{tabular}{lc}
\hline
LSA & Dynkin diagram and extended Dynkin diagram \\
\hline \\[5mm]
$A(m|n)$ & \\
         & \\[-15mm]
         & \mbox{\includegraphics{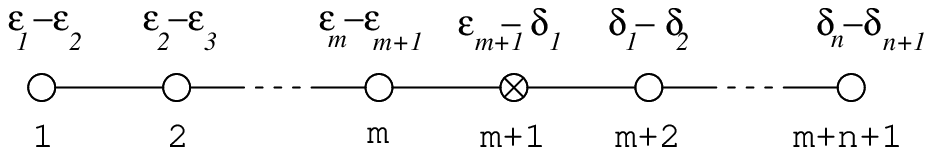}} \\[4mm]
         & \includegraphics{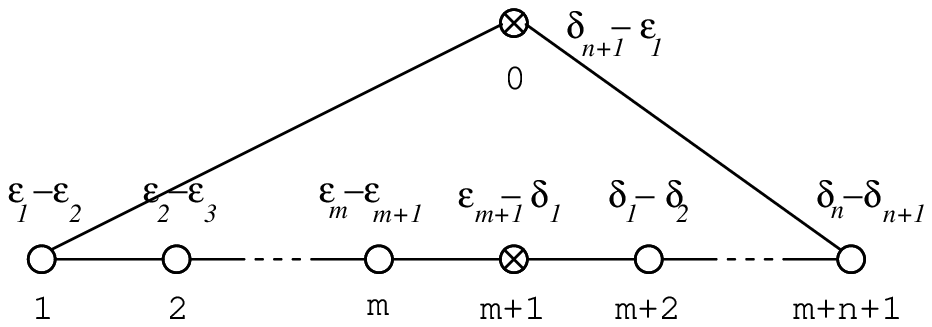}\\[10mm]
$B(m|n)$ & \\
         & \\[-15mm]
         & \mbox{\includegraphics{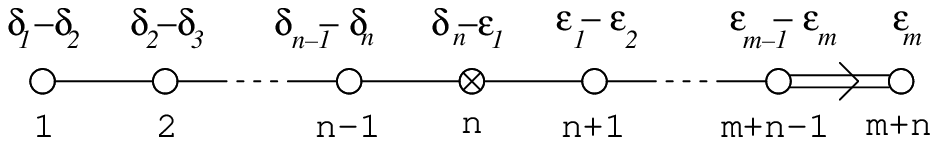}} \\[3mm]
         & \includegraphics{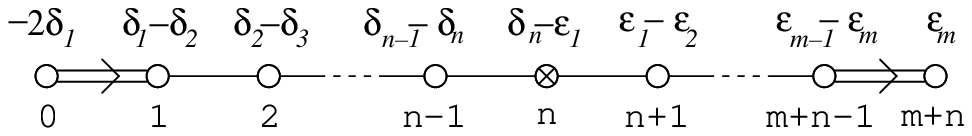}\\[10mm]
$B(0|n)$ & \\
         & \\[-10mm]
         & \mbox{\includegraphics{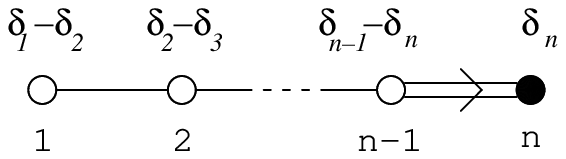}} \ \ \  \includegraphics{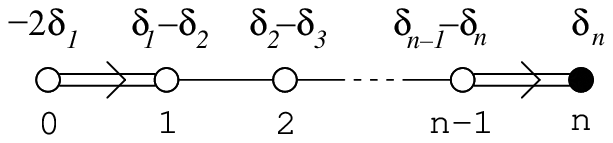}\\[10mm]
$D(m|n)$ & \\
         & \\[-15mm]
         & \mbox{\includegraphics{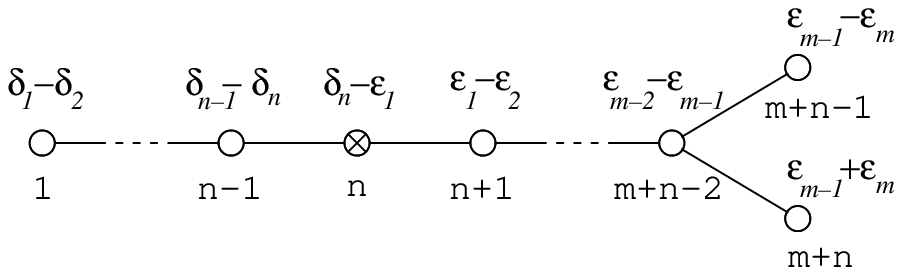}} \\[2mm]
         & \includegraphics{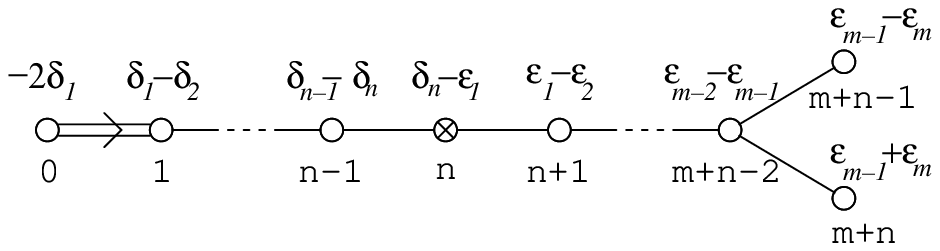}\\
$C(n)$ & \\
         & \\[-10mm]
         & \mbox{\includegraphics{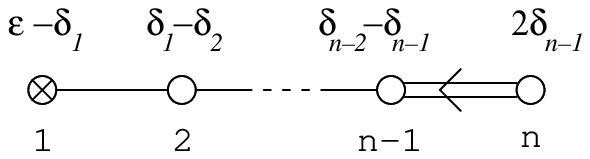}} \ \ \  \includegraphics{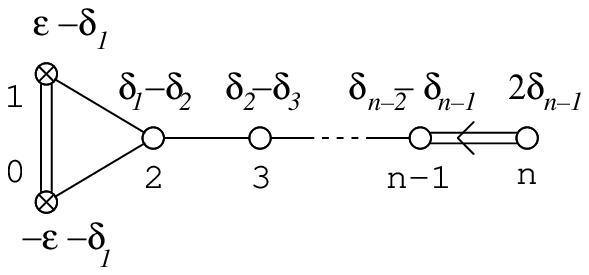}\\         
\hline
\end{tabular}

\newpage
\noindent
{\bf Table 2}.
Summary of the classification: all non-isomorphic GQS associated with a 
classical Lie superalgebra (LSA) are given. For each GQS, we list: the subalgebra $G_0$ 
(each $G_0$ contains the complete Cartan subalgebra $H$, so we only list the remaining part of 
$G_0=H+\cdots$); the length $\ell$ of the $Z$-grading; the number $N$ of annihilation operators;
the reference in the text where (an example of) $G_{-1}$ can be found.
\vskip 2mm
\noindent
{\small
\begin{tabular}{l|l|l|l|l}
\hline
LSA          & $G_0=H+\cdots$ & $\ell$ & $N$ & $G_{-1}$   \\
\hline \hline
$A(m|n)$  &  $sl(k|l)\oplus sl(p|q)$ &  3 & $(k+l)(p+q)$ &  (\ref{exA1}) or (\ref{exA2})\\
          & ($k+p=m+1$, $l+q=n+1$,  &    &  &  \\
          & \ $k+l\ne 0$, $p+q\ne 0$) &&&\\
          \cline{2-5}
          & $sl(k|l)\oplus sl(p|q) \oplus sl(r|s)$ & 5 & $(k+l)(p+q+r+s)$ & (\ref{A21}), (\ref{exA3}) \\
          & ($k+p+r=m+1$,  & 5 & $(p+q)(k+l+r+s)$ & (\ref{A22}), (\ref{exA4})\\
          & \ $l+q+s=n+1$, & 5 & $(r+s)(k+l+p+q)$ & (\ref{A23}), (\ref{exA5}) \\
          & \ $k+l\ne 0$, $p+q\ne 0$, $r+s\ne 0$) &&&\\
          \cline{2-5}
\hline  
$B(m|n)$ 
         & $sl(k|l)\oplus B(m-k|n-l)$ & 5 & $(k+l)(2m-2k+2n-2l+1)$ & (\ref{B21}), (\ref{B22}),  \\
         & ($k=0,\ldots,m$; $l=0,\ldots,n$; & & & (\ref{B7}), (\ref{B8}) \\
         & \ $(k,l)\not\in\{(0,0),(1,0)\} )$ &&& \\
         \cline{2-5}
         & $B(m-1|n) \qquad [(k,l)=(1,0)]$  & 3 & $2m+2n-1$ & (\ref{B5}) \\
\hline         
$B(0|n)$ & $sl(i)\oplus B(0|n-i)$ & 5 & $i(2n-2i+1)$ & (\ref{B021}) \\
         & ($i=1,\ldots,n$) & & & \\
\hline
$D(m|n)$ & $sl(k|l)\oplus D(m-k|n-l)$ & 5 & $2(k+l)(m+n-k-l)$ & (\ref{D10}), (\ref{D11}),  \\
         & ($k=0,1,\ldots,m$;  & & & (\ref{D21}), (\ref{D22}), \\
         & \ $l=0,1,\ldots,n$; &&& (\ref{D12}) \\
         & \ $(k,l)\not\in\{(0,0),(1,0),(m-1,n),(m,n)\} $) &&& \\
         \cline{2-5}
         & $D(m-1|n)\qquad [(k,l)=(1,0)]$  & 3 & $2(m+n-1)$ & (\ref{D7}), (\ref{D8}) \\
         \cline{2-5}
         & $sl(m|n) \qquad [(k,l)=[m,n)]$ & 3 & $(m+n)(m+n+1)/2-m$ & (\ref{D22a}), (\ref{D22b}) \\
         \cline{2-5}
         & $sl(m-1|n) \qquad [(k,l)=(m-1,n)]$ & 5 & $(m+n)(m+n+1)/2-m$ & (\ref{D42}), (\ref{D43}) \\
         \cline{2-5}
         & $sl(m-1|n)\qquad [(k,l)=(m-1,n)]$ & 5 & $2(m+n-1)$ & (\ref{D41}) \\
\hline         
$C(n)$   & $sl(k|l)\oplus D(1-k|n-1-l)$ & 5 & $2(k+l)(n-k-l)$ & (\ref{C2}), (\ref{C7}), \\
         & ($k=0,1$; $l=1,\ldots,n-2$) & & & (\ref{C9}), (\ref{C10}) \\
         \cline{2-5}
         & $C_{n-1}\qquad [(k,l)=(1,0)]$ & 3 & $2(n-1)$ & (\ref{C1}) \\
         \cline{2-5}
         & $sl(1|n-1)\qquad [(k,l)=(1,n-1)]$ & 3 & $n(n+1)/2-1$ & (\ref{C3}) \\
         \cline{2-5}
         & $sl(n-1)\qquad [(k,l)=(0,n-1)]$ & 5 & $n(n+1)/2-1$ & (\ref{c5.1}), (\ref{c5.3}) \\
         \cline{2-5}
         & $sl(n-1)\qquad [(k,l)=(0,n-1)]$ & 5 & $2(n-1)$ & (\ref{c5.2}) \\
         \cline{2-5}
\hline\hline         
\end{tabular}
}
\end{document}